\documentclass[aps,prapplied,twocolumn]{revtex4-2}
\usepackage{graphicx}
\usepackage{csquotes}
\graphicspath{ {./figures/} }
\usepackage{float}
\usepackage{amsmath}
\usepackage{amsfonts}
\usepackage{subfigure}
\usepackage{xcolor}

\usepackage[margin = 1in]{geometry}
\usepackage{multirow}
\usepackage{qcircuit}
\usepackage{braket}

\begin{document}

\title{Quantum-Processing-Assisted Classical Communications}
\author{Kelly Werker Smith$^{1,2}$ }
\thanks{kwsmith@fas.harvard.edu}
\author{Don Boroson$^{3}$}
\author{Saikat Guha$^{4}$}
\author{Johannes Borregaard$^{1}$}
\affiliation{$^1$Department of Physics, Harvard University, Cambridge, Massachusetts 02138, USA}
\affiliation{$^2$Harvard Quantum Initiative, Harvard University, Cambridge, Massachusetts 02138, USA}
\affiliation{$^3$Lincoln Laboratory, Massachusetts Institute of Technology, Lexington, Massachusetts 02421, USA}
\affiliation{$^4$Department of Electrical and Computer Engineering, University of Maryland, College Park, Maryland 20742, USA}
\date{\today}

\begin{abstract}
    We describe a general quantum receiver protocol that maps laser-light-modulated classical communications signals into quantum processors for decoding with quantum logic. The quantum logic enables joint quantum measurements over a codeword to  achieve the quantum limit of communications capacity. Our receiver design requires only logarithmically increasing qubit resources with the size of the codeword and accommodates practically relevant coherent-state modulation containing multiple photons per pulse. Focusing on classical-quantum polar codes, we outline the necessary quality of quantum operations and codeword lengths to demonstrate a quantum processing-enhanced communications rate surpassing that of any known classical optical receiver-decoder pair. Specifically, we show that a small quantum receiver of 4 qubits with operational errors of $\sim 0.2\%$ can already provide a $5$ percent gain in the communications rate in the weak signal limit. Additionally, we outline a possible hardware implementation of the receiver where efficient spin-photon interfaces such as cavity-coupled diamond color centers or atomic qubits are used to input the received photonic signal to a small scale quantum processor for decoding.  Our results outline a new, promising route for potential quantum advantage in classical communication with near-term, small-scale quantum computers.
\end{abstract}

\maketitle


\section{Introduction}

The ability to reliably transfer information over large distances is critical for a broad range of technologies ranging from the high-speed Internet to the global positioning system (GPS). Optical-frequency laser-light modulated communications affords far higher rates compared with its radio-frequency counterpart due to its greater spectral bandwidth and smaller diffraction spread. Nonetheless, optical communications is fraught with the need for precise pointing, acquisition and tracking (PAT) and with losses associated with absorption, scattering and turbulence in propagation through the atmosphere. However, free-space optical (FSO) satellite-to-satellite and deep-space links over vacuum are unfettered from the aforesaid limitations, making laser communications a very promising technology for high-rate links. Laser communications has therefore been pursued for missions such as the NASA/MIT Lincoln Laboratory Lunar Lasercom Demonstration (LLCD) program~\cite{Robinson2011}.

For the above-mentioned long-range FSO communications settings, stringent limitations on the transmitted power and on the aperture sizes of the transmitter-receiver telescopes severely restrict the achievable communications rate. Achieving the bandwidth limit imposed by the hardware constraints is a non-trivial task that involves careful optimization of the modulation format, error-correcting codes and receiver designs, along with efficient decoding algorithms that are capable of  countering the effects of transmission loss and operational errors~\cite{Hamkins2004,Cheng2006}. 

The maximum rate of information transfer is characterized by the channel capacity, which is the maximum number of bits that can be reliably communicated per use of the communication channel. The fundamental limit for classical information transfer is the Holevo bound~\cite{gordon, holevo73, levitin}, which provides an upper bound for the channel capacity. In particular, the Holevo bound exceeds the Shannon capacity associated with specific optical communication receivers~\cite{shannon}. In these cases, the Shannon capacity neglects the quantum mechanical properties of the communication channel and receiver measurement. This suggests that transmitter-receiver co-designs have to adopt a quantum mechanical description of the encoding and decoding of information to reach ultimate communication rates.   

Past work strongly suggests the necessity of entanglement within receivers~\cite{Chung_2017} to achieve information rates approaching the Holevo bound in the weak signal limit where the rate of received photons is small. This can be understood from the observation that collective measurements over multiple codeword symbols can result in a superadditive gain over individual symbol detection~\cite{Guha2011}. For strongly attenuated signals, only a few photons are received across many symbols and the collective measurements become projections onto entangled photonic subspaces. Recently, superadditivity in this weak signal regime has been demonstrated with trapped ions~\cite{Delaney2022} and optical interferometers~\cite{cui} but none of the demonstrated designs have any guarantees of reaching the Holevo bound for larger, more practically relevant, code sizes. For example, the LLCD used a codeword length of 15,120 bits \cite{moision, boroson}.

Classical-quantum polar codes are known to asymptotically approach the Holevo bound for the pure-loss optical channel, with the aid of an entangling quantum successive cancellation (SC) decoder~\cite{saikat-wilde1}. However, the practical implementation of the required measurements, performance-evaluation in the non-asymptotic limit, and robustness to operational errors in the quantum decoding hardware has so far not been carefully addressed. As such, it is unknown whether or not practically feasible receivers can be realized for finite-size codes.     

In this work, we develop a general framework for mapping photonic states used for optical communications to quantum processors for subsequent decoding with quantum logic. We achieve this utilizing photonic-to-binary conversion techniques~\cite{emil,emil2}, which we extend to the practically relevant multiple-photon regime. Once the received state is loaded into a processor, quantum gates can be used to implement the specific measurements of the decoder in a programmable way. To illustrate our framework, we design a concrete quantum receiver capable of performing SC decoding of classical-quantum polar codes with only logarithmically increasing qubit resources with the codeword length. 

\begin{figure*}[t!] 
    \centering
       \includegraphics[height=15cm, width=15cm]{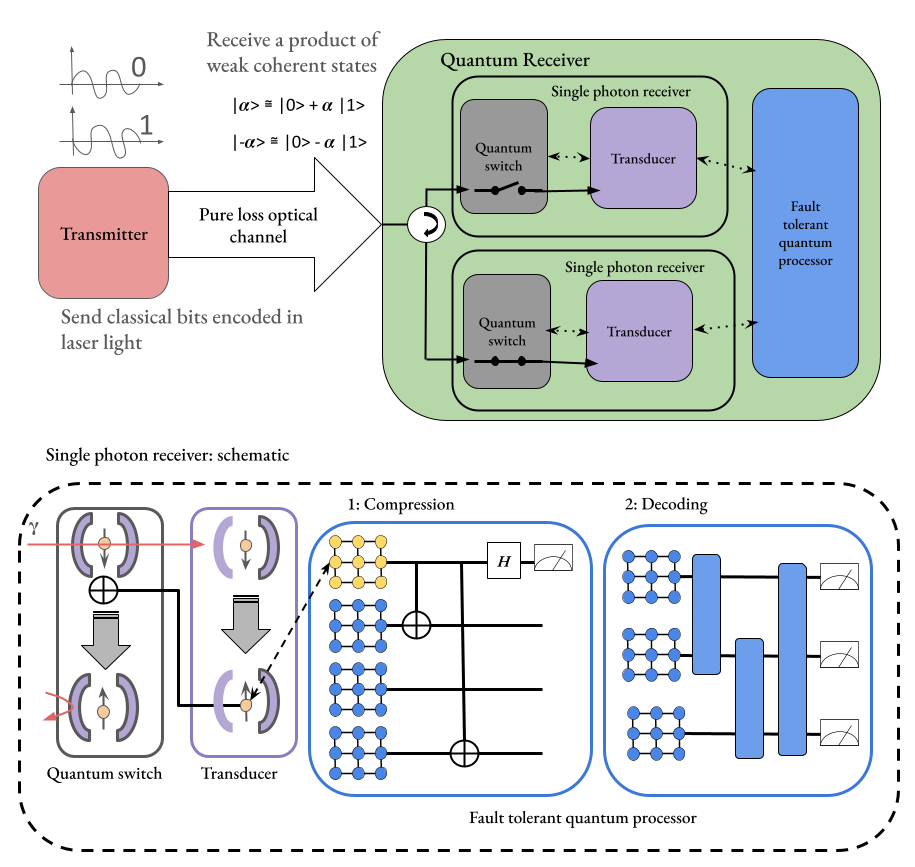}
        \caption{Schematic representation of the full communication setup including the transmitter, channel and quantum receiver. The receiver, which is the focus of this work, has a modular design to accommodate multiple photon inputs. In each time-bin, the signal is routed to the quantum switches, which will transmit (reflect) the signal if the corresponding memory bank in the quantum processor is empty (full). Single photon receiver: A photon ($\gamma$) arriving in the first time-bin is directed to the transducer qubit. A CNOT gate between the transducer qubit and the quantum switch qubit flips the switch from transmission to reflection. The state of the transducer qubit is mapped to a logical control qubit, and a series of logical CNOT gates flip the processor qubits to encode the photon arrival bin and phase information (the compression step). Next, the control qubit is decoupled from the target qubits with a measurement in the X basis. Finally, gates and measurements are performed on the remaining target qubits to decode the received classical bit string (the decoding step).}
    \label{fig:rxSchematic}
\end{figure*}

Our design consists of a photon-to-qubit transducer, which maps the photonic amplitude information into a small-scale, fault-tolerant quantum processor through state injection (see Fig.~\ref{fig:rxSchematic}). The transducer can be implemented with atomic or atom-like qubits strongly coupled to optical cavities, in a manner similar to recent experimental demonstrations of quantum assisted non-local interferometry~\cite{stas2025}. We quantitatively investigate the performance of the receiver for finite codeword lengths in the presence of realistic imperfections such as quantum gate noise. We show that a superadditive rate improvement of 5\% over any known classical receiver can be demonstrated with modest sized quantum processors of 4 qubits, with qubit error rates of a few permille in the weak signal limit where each symbol contains an average of $\sim0.001$ photons. Compared to previous designs of quantum processor-assisted optical communications systems~\cite{Crossman2023}, our design targets practically relevant codes with rigorous guarantees of achieving the Holevo bound in the asymptotic limit, does not require any optical-to-microwave transduction, and has known decoding circuits. 

Our analysis suggests that small-scale quantum error-corrected devices could be used to demonstrate an advantage over classical techniques in optical communications. While further development is needed to make quantum-enhanced classical communications of practical use, our work opens up several promising new directions for future research in this field.

\section{Receiver design}\label{rx_design}

As shown schematically in Figure \ref{fig:rxSchematic}, we model the communication channel as a pure-loss optical channel, which forms a building block for modeling optical communications over diffraction-limited free-space links~\cite{Shapiro2005}. This modeling choice is motivated by the space-based free-space communication scenarios of primary interest in this work, where diffraction and finite aperture sizes dominate. In such settings, phase noise accumulated during propagation is negligible compared to photon loss, and modeling the channel as pure loss is a  justified approximation~\cite{Shapiro2005, giovannetti}. We focus on the encoding of classical information in quantum polar codes following the approach of Ref.~\cite{saikat-wilde2}, but our receiver design is generally applicable to any encoding across multiple optical pulses in the limit where the probability of simultaneous arrival of multiple photons is negligible. 

\subsection{Signal encoding}
The classical information bits are first translated into multi-bit classical codewords of the polar code. Polar coding exploits \emph{channel polarization} to ensure that all the channels are either near useless (bad) or near perfect (good). We can then send our information bits through the good channels and fix the bits input to the bad channels to some value known by both sender and receiver (so-called \enquote{frozen} bits). For more details on polar codes, we refer the reader to Ref. \cite{arikan}. The codewords are then encoded bit-by-bit into the phases of individual optical pulses using the binary phase shift keying (BPSK) protocol. We model this as encoding bit 0 (1) as an optical coherent state $\ket{\alpha}$ ($\ket{-\alpha}$). The mean photon number of the state is given by $\overline{n} = \alpha^2$. We can take $\alpha$ to be real for the BPSK alphabet without loss of generality ~\cite{saikat-wilde2}. The encoding happens sequentially such that an $N$-bit long codeword is encoded into a temporal sequence of $N$ consecutive coherent states. The ratio $\frac{K}{N}$ between the number of information bits $K$ and the length of the codewords $N$ is referred to as the code rate.   

As an illustrative example, we consider $N=4$ and $K=2$. With two frozen bits, the information bits can take on the values 00, 01, 10, and 11 and our corresponding polar-coded codewords $c=c_1c_2c_3c_4$ are 0000, 1111, 1010 and 0101, respectively~\cite{arikan}. The encoded states are thus of the form $\ket{\psi_{c}}=\bigotimes_{i=1}^4\ket{(-1)^{c_i}\alpha}$, with coherent state $\ket{\alpha} = e^{-\frac{1}{2}|\alpha|^2} \sum_{j = 0}^\infty \frac{\alpha^j}{\sqrt{j!}} \ket{j} = e^{-\frac{1}{2}|\alpha|^2}[\ket{0} + \alpha \ket{1} + \frac{\alpha^2}{\sqrt{2} }\ket{2} + \dots ]$. Note that here and in what follows, $\ket{0}$ and $\ket{1}$ denote the Fock number state, not the classical binary digit. In particular, each digit in a ket represents a time-bin and the numeral is the number of photons in that time-bin. $\ket{\psi_c}$ can be expanded in the weak field limit as $\ket{\psi_{c}}\approx e^{-\frac{1}{2}|\alpha|^2}\bigl[\ket{0000}+\alpha\ket{\psi^{(1)}_{c}}+\alpha^2\ket{\psi^{(2)}_{c}}+\frac{1}{\sqrt{2}}\alpha^2\ket{\psi'^{(2)}_{c}}+O(\alpha^3)\bigr]$ where

\begin{eqnarray}
\ket{\psi^{(1)}_{c}} &=&(-1)^{c_1}\ket{1000} +  (-1)^{c_2}\ket{0100} \nonumber \\
&&+(-1)^{c_3}\ket{0010} +  (-1)^{c_4}\ket{0001}] \\
\ket{\psi^{(2)}_{c}} &=& [ (-1)^{c_1+c_2}\ket{1100}+(-1)^{c_1+c_3}\ket{1001}\nonumber \\
&&+(-1)^{c_1+c_4}\ket{1001}+(-1)^{c_2+c_3}\ket{0110} \nonumber \\
&&+(-1)^{c_2+c_4}\ket{0101}+(-1)^{c_3+c_4}\ket{0011}] \\
\ket{\psi'^{(2)}_{c}}&=&[\ket{2000}+\ket{0200}+\ket{0020}+\ket{0002}]. \qquad
\end{eqnarray}
We expand in the weak-field limit since this a regime where quantum, collective decoding receivers could be beneficial over a classical counterpart. However, the regime for quantum assistance is not demarcated by a sharp line because there are many tradeoffs to be considered in the design of practical communication systems. For example, to achieve a particular average power efficiency, the designer could adjust the output power of the transmitter amplifier, the area of the transmit telescope, or the area of the receive telescope. In addition, there might be constraints on bandwidth, peak power or cost. While Holevo is the ultimate capacity limit, practical systems might meet the desired receiver specifications operating below Holevo where classical techniques can reach the required rates. Further discussion of the utility of collective decoding versus classical techniques can be found in Refs. ~\cite{giovannetti, dolinarUlt, borson_book}

\subsection{Qubit mapping}
The first task of the receiver is to map the photonic state into a quantum processor. The mapping is facilitated by a \emph{transducer} qubit. The transduction from photonic to matter-based encoding can be implemented through a stimulated Raman adiabatic passage as theoretically described in Refs.~\cite{cirac,kollath2024, huang} and experimentally demonstrated with cavity-coupled single atoms in Refs.~\cite{ritter,boozer}. In these schemes, the cavity-coupled transducer qubit is encoded in two stable ground levels of a three level spin system. The qubit is initially prepared in ground state $\ket{\downarrow}$, which is coupled to an excited state through a cavity enhanced optical transition. The absorption of a photon excites the spin system and a control laser pulse adiabatically transfers the population through the excited state to the other ground state $\ket{\uparrow}$. This amounts to the transformation: $\ket{1}\ket{\downarrow}\to\ket{0}\ket{\uparrow}$, which corresponds to the coherent absorption of the photon into the transducer qubit. Through careful design of the incoming photonic pulses and laser control fields, near unit efficiency can be achieved~\cite{cirac,kollath2024}.   

The state of the transducer qubit is then transferred to a logical qubit in the quantum processor through qubit injection~\cite{luo, gidney, lao, Li_2015}. At the abstract level, this requires the ability to perform qubit stabilizer measurements between the transducer qubit and a subset of the physical qubits of the logical code. Depending on the specific hardware implementation, this can be achieved in several ways. For atomic based systems, the possibility to physically transfer atoms between a cavity region and a processing region~\cite{covey2023,Tamara2021} would allow for direct injection of the transducer qubit. For solid state hardware~\cite{stas2025}, pre-shared entanglement stored in nearby nuclear spin qubits could be used to teleport the information into the processor, which could consist of the same~\cite{singh2024} or different hardware~\cite{mokeev2025}.     

Once the state of the transducer qubit has been transferred to a logical qubit in the quantum processor, the processor records the photon arrival bin and phase information using the binary encoding approach outlined in \cite{emil, emil2}. Specifically, the logical qubit that stores the transducer state serves as a control qubit for a series of logical CNOT gates. The vacuum ($\ket{0000}$) and single photon component ($\ket{\psi^{(1)}_{c}}$) of $\ket{\psi_c}$ can be mapped to the processor by changing the target qubits of these logical CNOT gates depending on the time-bin (the \enquote{Compression} step in Figure \ref{fig:rxSchematic}). For a codeword of length $N=4$, $\lceil \log_2(4+1) \rceil = 3 $ logical qubits are required, while for a codeword of length $N$, $\lceil \log_2(N+1) \rceil$ logical qubits would be required ~\cite{emil,emil2}. This is an exponential reduction in qubit resources compared to a unary encoding approach where there is a qubit for each time-bin.

 However, there are a couple of caveats. First, the logical control qubit has to be decoupled from the rest of the qubit register to complete the mapping. As outlined in Refs.~\cite{emil,emil2}, this can be done by a final $X$-basis measurement, projecting onto the states $\ket{\pm}=(\ket{0}\pm\ket{1})/\sqrt{2}$ following the CNOT encoding sequence of each time-bin. Depending on the measurement result, $x_i \in \{0,1\}$, a phase correction should be applied to the qubit state.  These corrections can be recorded and applied later in the classical decoding of the information. 

Second, and more problematic, the protocol only allows for the mapping of the vacuum and single photon components of $\ket{\psi_c}$ since the two-photon terms would lead to errors in the binary encoding of the time-bin information. We solve this issue by adopting a modular design where the receiver consists of multiple logarithmically sized single photon sub-receivers (see Figure~\ref{fig:rxSchematic}). The key element of our multiple-photon receiver is the addition of a qubit switch, which routes the incoming photons between the single photon qubit registers. The switch can be implemented using as second cavity-coupled spin system which controls the input field to the transducer qubit. In particular, if only one of the spin states has a cavity-resonant optical transition and the cavity is a balanced two-sided cavity, then the cavity-qubit system will be reflecting (transmitting) if the qubit is in the coupled (uncoupled) state~\cite{sorenson}. Without loss of generality, we can assume that if the switch qubit is in state $\ket{\downarrow}$ ($\ket{\uparrow}$), the photon is transmitted through (reflected from) the cavity. 

The quantum switches are initialized in $\ket{\downarrow}$ such that an incident photon is transmitted to the corresponding single photon receivers. The incoming photons will always be routed from the first to the last switch-register module regardless of the time-bin. The single photon receiver protocol as described above proceeds within each register with the addition of an extra CNOT gate between the switch qubit and transducer qubit before state injection. The transducer qubit is the control and the switch qubit is the target of the CNOT. If a photon was absorbed by the transducer qubit in that time-bin, the CNOT will change the qubit switch from transmissive to reflective. This ensures that when one photon is stored in a quantum register module, any subsequent photons will be stored in the next registers. The CNOT interaction between the transducer qubit and the switch qubit can be implemented in a similar way as the injection of the transducer qubit into the processer via either direct or entanglement-assisted gates depending on the hardware implementation. 

Using $k$ modules our receiver is thus able to map $k$ photons into $k\log{(N+1)}$ logical qubits. The number of physical qubits required per logical qubit depends on the error rate and code used. For example, quantum low-density parity check codes have been designed that can achieve a physical to logical ratio of $\sim$10 to 1 \cite{xu2025}. For the transducer and quantum switch, one physical qubit is required for the transducer and one physical qubit is required for the switch, with one transducer and one switch per single photon receiver. 

Note that our scheme as described so far cannot accommodate multiple photons arriving in the same time-bin since our qubit switch can only interact with a single photon at a time. In other words, we are able to map all leading terms of $\ket{\psi_c}$ except the $\ket{\psi'^{2}_c}$ term similar to the unary approach. However, in the weak signal limit, the probability of simultaneously arriving photons is suppressed by a factor of $\frac{1}{N}$ compared to the probability of two photons arriving in different time-bins ensuring that such error is suppressed as the codeword length increases. In general, for fixed $k \ll N$, terms with simultaneously arriving photons are suppressed with $N$ relative to the single photon per time-bin terms. 

For strong signals, the use of spatial splitting can separate simultaneously arriving photons. This breaks the received signal up into multiple channels that are each in the weak signal limit discussed above. The states incident to the receiver can be separated using 50/50 beam splitters. To split the input into $m$ spatial modes, $m-1$ beam splitters are required. Each output mode would be routed to a single photon receiver, requiring $m$ single photon receivers. This approach comes at the cost of more qubits and more complicated decoding, as the projectors and measurements must now be determined across multiple receivers.

Our receiver provides a logarithmic scaling for the number of qubits with $N$, as opposed to the linear scaling of a unary encoding. The scaling factor will depend linearly on the number of expected photons received ($k$) and the number of physical qubits per logical qubit. The scaling difference with the unary encoding is particularly important in the weak field regime, where the mean photon number per time-bin is low such that only a few received photons are expected over the entire codeword. Considering a codeword of length $N$, and an expected mean photon number per time-bin  at the receiver of $\alpha^2$, our receiver would require $~N\alpha^2\log{(N+1)}$ qubits while the unary encoding receiver would always require $N$ qubits. The values of $\alpha$ received for a practical system depend on tradeoffs with peak power, band width, aperture size and cost as mentioned in Section II A. If a system is designed such that $N\sim100$ and $\alpha^2\sim0.01$, our receiver offers an order of magnitude reduction of the qubit resources. However, in other regimes, the unary encoding might be more favorable.

\subsection{Successive cancellation decoding}

Once the photonic states are mapped to the quantum processor, quantum gates can be used to decode the stored information (see the \enquote{Decoding} step in Figure \ref{fig:rxSchematic}). Finding the optimal decoding strategy is, in general, a non-trivial task. Yet for polar codes, it was shown in Ref.~\cite{saikat-wilde1} that quantum successive cancellation decoding can saturate the Holevo bound in the asymptotic limit. It is not known how to implement such a decoding with direct photon detection.  However, once the photonic state is mapped into qubit registers, quantum logic operations can be used to perform the decoding. We show a general procedure for implementing this decoder with quantum logic following the photonic-to-qubit mapping in our quantum receiver. We detail the decoding scheme for $N = 4$ and $N = 8$. While the procedure for decoding higher $N$ is the same, we note that there is no general formulation for the polar code construction for abitrary $N$. This is because the polarizing transformation is not step-wise and isotropic \cite{arikanlecture} with the channels it acts on. However, efficient constructions for good classical polar codes for higher $N$ have been devised~\cite{Tal2013}, which are likely to be generalizable for quantum polar codes.

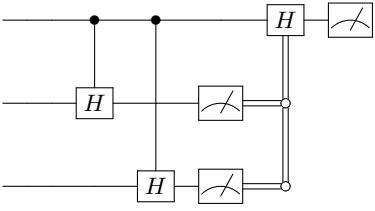
\begin{figure}
\[ \Qcircuit @C=1em @R=2em {
& \qw & \qw  & \ctrl{1} & \ctrl{2} & \qw & \gate{H} & \meter\\
& \qw & \qw & \gate{H} & \qw & \meter & \controlo \cw  \cwx \\
&  \qw & \qw & \qw &\gate{H} & \meter & \controlo \cw  \cwx
}\]
\caption{Decoding circuit for $N=4$. The double wires with open circles represent classical control conditioned on a measurement outcome of 0. The $H$ denotes a Hadamard gate.}
\label{fig:decoding1}
\end{figure}

\begin{figure}
    \includegraphics[height=5cm, width=9cm]{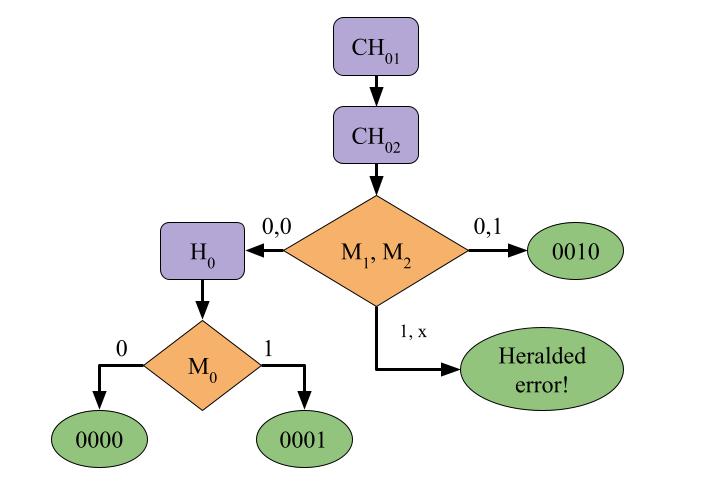}
    \caption{A decision tree to decode the message for $N=4$. $M_i$ represents a measurement of qubit $i$ in the $Z$ basis. The binary labels denote measurement outcomes. $CH_{ij}$ is a controlled Hadamard gate with qubit $i$ as control and $j$ as target. $H_i$ is a Hadamard gate on qubit $i$. }
    \label{fig:decoding2}
\end{figure}

The general idea behind the SC decoding is to perform successive joint measurements on the processor qubits. Each measurement decodes a single bit, and the next measurement is conditioned on the results of the preceding ones. Each bit decision eliminates half of the (remaining) codewords. The measurements are, in general, projections onto multi-qubit entangled states. We provide the specific projections for $N=4$ and $N=8$ in the supplementary material \cite{supplement}. The general decoding circuit and decision tree for $N=4$ is shown in Figs.~\ref{fig:decoding1}-\ref{fig:decoding2} 

As an illustrative example, we consider the decoding of the specific codeword of 0001 ($N=4$). For simplicity, we restrict ourselves to the vacuum and single photon subspace where the initial qubit state can be expressed as (with normalization factor $\frac{1}{1+4\alpha^2}$):
\begin{equation}
    \ket{\uparrow \uparrow \uparrow} - \alpha [  \ket{\downarrow \uparrow \uparrow} +\ket{\downarrow \uparrow \downarrow}  
    +\ket{\downarrow \downarrow \uparrow}+\ket{\downarrow \downarrow \downarrow}]
\end{equation}
The two controlled Hadamard gates (see Fig.~\ref{fig:decoding1}) transform the state to
\begin{equation}
   \ket{\uparrow \uparrow \uparrow} - 2 \alpha \ket{\downarrow \uparrow \uparrow}  \\
\end{equation}

The second and third qubits are now measured in the Z-basis, with both measurements yielding the outcome \enquote{$\ket{\uparrow}$}.   Following the decision tree, an $H$ gate is then applied to the first qubit resulting in the state:
\begin{equation}
\left(\frac{1}{\sqrt{2}} -\sqrt{2} \alpha\right)\ket{\uparrow} + \left(\frac{1}{\sqrt{2}} +\sqrt{2}\alpha\right)\ket{\downarrow}.
\end{equation}
The probability of measuring \enquote{$\ket{\downarrow}$} and correctly decoding 0001 is therefore $\frac{2\alpha^2 + 2\alpha + 0.5}{4\alpha^2 + 1}$. Notably, if the outcome \enquote{$\ket{\uparrow}$} is obtained, the decoding is unsuccessful: the receiver decides that 0000 was sent rather than 0001. As the length of the codeword increases, the decision error probability becomes very small as shown in Ref. \cite{saikat-wilde2}. To maximize the communication rate, codewords are assigned according to the relative likelihood (the input distribution) of the input strings. In practical scenarios, the codewords might encode symbols, for example letters in the English language. Since \enquote{e} is the most frequent letter in English, the codeword with the smallest decision error would be assigned to \enquote{e}. To calculate the maximum achievable rate, or capacity, it is necessary to optimize over the input distribution of codewords. We refer the reader to Section III below and the supplementary material for further details on this optimization \cite{supplement}. Table 1 shows a decoding table that can be used to decide on the transmitted message given the measurement results.

\begin{table}
    \caption{Decoding table for $N=4, K=2$. The optimal distribution only has 3 codewords with nonzero input probability. The optimal distribution is given in the supplementary material \cite{supplement}}
    \begin{tabular}{ |p{0.5cm}|p{0.5cm}|p{0.5cm}|p{0.75cm}|}
     \hline
     Qubit 0 & Qubit 1 & Qubit 2 & Decision  \\
     \hline
     $0$ & $0$ &  $0$  &   0000 \\
     $1$ & $0$  &  $0$  &   0001  \\
     $0,1$ & $0$  &  $1$  &   0010 \\
     $0,1$ & $1$  &  $0,1$  &   error \\
     \hline
    \end{tabular}
\end{table}

The depth of the circuit for the compression step scales as $\sim N\log_2{N}$. In the decoding step, there are $\sim \log_2{N}$ input qubits. An arbitrary $q$ qubit unitary can be implemented by a circuit depth of $O(4^{q})$ \cite{shende}, which in our case with $q=\log_2{N}$, is $O(N^2)$. Even if the decoding requires processing and measuring out each qubit individually, the depth is upper bounded by $\sim N^2\log_2{N}$ in the worst case. Overall, the processing circuit scales as $O(N^2\log_2{N})$ in the worst case.

\section{Receiver performance}

From the classical-quantum polar code construction in \cite{saikat-wilde1}, we determined the SC decoding projectors for $N=2, 4, 8$ and code rates of $\frac{1}{2}$. From these projectors, we calculated the photon information efficiency (PIE) for the code and receiver scheme. PIE, in bits/photon, is the average number of bits of information that can be reliably transmitted in each photon and is calculated by dividing the channel information rate\footnote{Note that we find a \emph{rate} in general less than the ultimate capacity because we are assuming the use of the quantum SC decoder described in \cite{saikat-wilde1}. To find the capacity, we would have to optimize over all possible measurements.} by the mean photon number of the received pulse. We focus on small code lengths and the weak signal limit ($\alpha \ll 1$) to investigate the onset and robustness of superadditivity.  

The projectors used in the performance analysis are designed for the vacuum and single photon subspace. Nonetheless, we do include the multiple-photon contributions assuming that they lead to non-informative measurement outcomes corresponding to a uniform guess probability of the input message. While this design of the quantum receiver is sub-optimal, we do not expect any significant degrading of the performance in the weak field limit (because the probability of higher photon number events is relatively small) while it significantly relaxes the quantum resources for near-term demonstrations.

\subsection{Performance without errors}

\begin{figure}
    \includegraphics[height=6cm, width=8cm]{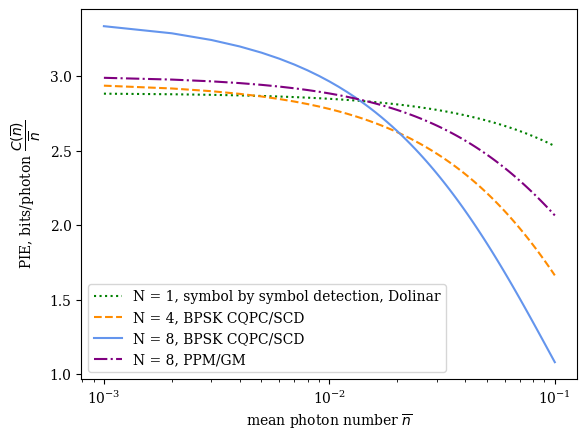}
    \caption{PIE versus mean photon number for CQ polar codes, Dolinar symbol by symbol detection and PPM.}
    \label{fig:perfNoErr}
\end{figure}

The PIE for polar codes compared to the best known classical symbol-by-symbol detection with a Dolinar receiver~\cite{dolinar} and joint detection Green Machine receiver~\cite{cui} is plotted versus mean photon number ($\bar{n}$) per time-bin in Figure~\ref{fig:perfNoErr}. Already at $N=4$, we see the onset of superadditivity where our quantum receiver outperforms the Dolinear receiver for $\bar{n}\lesssim 0.4\%$. Furthermore, at $N=8$, our quantum receiver surpasses the best performance of the Green Machine receiver for a comparable 8-ary pulse position modulation (PPM) encoding~\cite{cui} for $\bar{n}\lesssim 1\%$. This shows that even for finite-length codewords, our quantum receiver is able to surpass the performance of the best known classical decoders in the weak field limit. As the codeword length increases, larger gains are expected but quantifying this requires the identification of larger polar codes, which is beyond the scope of this work. However, we note that Ref.~\cite{saikat-wilde1} shows that the ultimate Holevo limit is achievable in the asymptotic limit of infinitely long codewords and the SC decoder.  

\subsection{Performance with errors}

Our receiver design has two distinct elements: the photon-to-qubit transducer and the quantum processor. Notably, the former acts on the physical qubit level where errors from imperfect photon absorption, switch operation, and injection directly degrade the information received from the photonic signal. On the other hand, the latter involves processing the information fault-tolerantly through encoded logical qubits. 

To investigate the potential for demonstrations of quantum processing enhanced classical communication with near-term quantum hardware, we conceptually distinguish between the physical error rate in the transduction step and the logical error rate of the processor. In our numerical investigations, we perform an optimization of both the input distribution and $\alpha$ to find the maximum bits/photon for a given error rate. \footnote{This was necessary since the bits/photon is no longer monotonic at low enough $\alpha$. The ability to distinguish the codeword states scales with $\alpha$, but the error probability does not. So, at low enough $\alpha$, the distinguishing signal drops below the noise floor set by the error rate and the PIE no longer increases with decreasing $\overline{n} = |\alpha|^2$.} 

It is sufficient for the purposes of this analysis to consider an input pure state $\rho = \ket{\psi}\bra{\psi}$, with $\ket{\psi} \in \{\text{polar codewords} \}$, because the pure-loss optical channel with BPSK has pure state outputs~\cite{saikat-wilde1}. The model of the transmission channel as a pure-loss optical channel is an acceptable approximation since photon loss, for instance from diffraction in free space propagation, is the dominant source of error for the transmission settings relevant to this work \cite{giovannetti, Shapiro2005}, as mentioned in Section \ref{rx_design}.

\begin{figure}
    \centering
    \subfigure[]{
        \includegraphics[height=6cm, width=8cm]{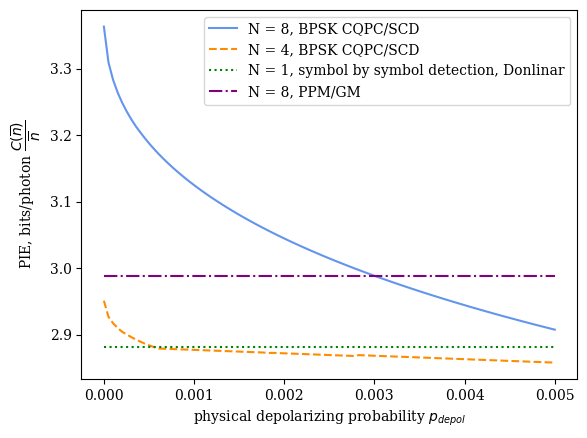}
        }
    \subfigure[]{
        \includegraphics[height=6cm, width=8cm]{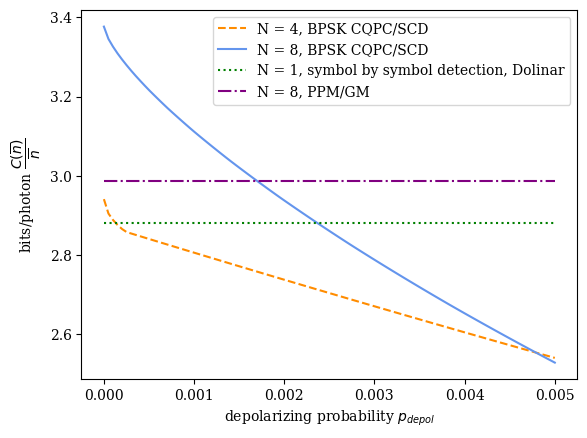}
    }
    \caption{ Receiver performance with (a) transducer errors and (b) gate errors in the compression and decoding steps. The green dotted line denotes the PIE with ideal symbol by symbol detection. In (a), we modeled the errors with depolarizing noise at the physical level on the input photonic state. In (b), we used a total depolarizing channel (unbiased Pauli channel) at the logical level for all interactions.}
    \label{fig:perfErr}
\end{figure}

\textbf{Transducer errors:} Before the input state is mapped to a fault-tolerant quantum processor, there can be unprotected errors at the physical level in the transducer step. These errors can come from imperfect photon absorption, state injection and switch operation. The error rates of each of these will depend on the quality of the spin-cavity interfaces~\cite{cirac,kollath2024,sorenson}, the quality of the CNOT operations between the transducer qubit and the switch qubit as well as general state preparation errors and measurement errors. Detailed error modelling of each of these elements can be peformed for specific hardware implementations but to obtain a general estimate of the required error level for demonstrating quantum advantage, we model the effect of these errors as a general depolarizing noise on the injected state in the processor. 

By directly injecting the input photonic state into an error-corrected matter qubit, the subsequent time-bin encoding procedure and final decoupling measurement are performed at the level of matter qubits rather than photons, and are protected by the error-correcting code. As such, photonic loss can only occur prior to absorption or during the physical absorption process itself. Loss prior to absorption is indistinguishable from transmission loss and acts to reduce the signal amplitude, while loss during absorption manifests as spontaneous emission. Treating this process as depolarizing noise provides a conservative lower bound on the receiver performance. The noise model can be modified to account for the biases of specific hardware platforms.

The effects of transducer errors on the PIE are plotted in Figure \ref{fig:perfErr}(a). We see that for physical depolarizing probabilities  of up to 0.3\%, superadditivity is present. Additionally, as $N$ increases from 4 to 8, we see a change in scaling, with a more rapid decrease in the PIE. Nonetheless, the superadditive gain sill increases for a fixed depolarizing probability showing that as the codeword increases, higher error rates can be tolerated while maintaining a quantum advantage. A qualitative explanation is that the capacity increases super-additively with the codeword length ($N$), whereas the effective depolarizing error rate of the transduction step grows only approximately linearly. This occurs because each time-bin in the codeword, whether it contains a photon or not, undergoes a physical transduction operation, leading to a linear scaling of the effective error rate with $N$, provided that the per-operation error remains $\ll 1/N$. 

\textbf{Logical errors:} We model the logical quantum gate errors as general Pauli noise channels acting at the logical level on each qubit participating in the gate operation. The encoding and decoding circuit with errors for $N=4$ is modeled as depicted in Figure~\ref{fig:errorcircuit}. Errors in the compression step are represented by the block labeled $\mathcal{E}$. The faulty binary encoding is modeled as a two-qubit Pauli noise channel on the control qubit that stores the photon arrival information and the target processor qubit after each CNOT interaction. Meanwhile, errors in the photon decoupling operation are modeled with single-qubit Pauli noise channels before the measurement. Further details on the error modeling in the compression step is presented in the supplementary material \cite{supplement}. 

In the decoding step, each two-qubit gate is followed by a two-qubit Pauli noise channel ($\mathcal{N}$ block) while single-qubit gates are followed by a single qubit Pauli noise channel ($\mathcal{M}$ block). In Figure~\ref{fig:perfErr}(b), we plot the performance considering that each error channel consist of independent single-qubit depolarizing channels with error probability $p_{depol}$. Results for some heuristic correlated error channels are presented in the supplementary material but we found little variation in the performance under the different error models \cite{supplement}. We show the results for unbiased Pauli noise channels as a conservative lower bound on the receiver performance, but the model can be modified to account for biased noise depending on the hardware implementation.

\begin{figure}
\[ \Qcircuit @C=1em @R=2em {
& \qw & \multigate{2}{\mathcal{E}}  & \ctrl{1} & \multigate{1}{\mathcal{N}} & \ctrl{2} & \gate{\mathcal{N}} & \qw  &\gate{H} & \gate{\mathcal{M}} & \meter\\
& \qw & \ghost{\mathcal{E}} & \gate{H} & \ghost{\mathcal{N}} & \qw & \qw & \meter & \controlo \cw  \cwx \\
&  \qw & \ghost{\mathcal{E}} & \qw & \qw &\gate{H} & \gate{\mathcal{N}} &  \meter & \controlo \cw  \cwx
}\]
\caption{The $N = 4$ decoding circuit with errors. The faulty photon-to-qubit mapping in the compression step is represented by the block labeled $\mathcal{E}$ and includes errors in the time-bin binary encoding interactions and in the photon decoupling. Each two-qubit gate in the decoding step is followed by a two-qubit Pauli noise channel ($\mathcal{N}$ block) while single-qubit gates are followed by a single qubit Pauli noise channel ($\mathcal{M}$ block).}
\label{fig:errorcircuit}
\end{figure}

Superadditivity is no longer achieved once errors exceed the permille level for the noise model considered here but similarly to the transducer errors, we observe that the performance still increases with $N$ for a fixed error probability, which we assign to the interplay between the super-additive gain and the roughly linear increase in overall error probability with $N$.

Performing a numerical simulation with transduction and logical errors, we find that a quantum receiver with 4 logical qubits and a physical error in the transducer step of $\sim 0.3\%$ or with quantum gate errors of $\sim 0.2\%$ can provide a 5 percent gain in communication rate in the weak signal limit. Experimentally achieved error rates for spin-photon cavity systems are at the percent level and have clear paths to improve performance~\cite{bhaskar, stas, Knaut}. Additionally, physical qubit-gate errors below 0.1\% has been demonstrated with trapped ion processors~\cite{hughes2025} indicating that first demonstrations of quantum advantage might not require logical encoding.      

\section{Conclusion and outlook}

In conclusion, we have designed and analyzed a receiver that makes an entangling measurement over multiple modulated laser-light pulses, and achieves superadditive communications capacity better than any known classical receiver, and in-principle capable of attaining the Holevo limit of capacity, using the quantum polar code and a quantum processor. Our receiver protocol provides a general framework in which the multi-photon manifold of coherent-state-modulated optical communications can be mapped effectively to matter qubits. Quantum gates are then performed on the matter qubits to decode the received message. To our knowledge, this is the first superadditive-capacity receiver design with a sub-exponential scaling in the number of its operations with codeword size~\cite{wilde}, and to concretely address how to transfer coherent optical signals into matter qubits in order to utilize a discrete variable paradigm of quantum information processing~\cite{daSilva}. 

Our receiver consists of two stages: a physical transduction stage and a logical processing stage. We have outlined how the transducer can be implemented with efficient spin-photon interfaces such as atoms and solid-state defect systems in optical cavities. Through optically mediated quantum teleportation, the transferred information can then be injected into a quantum processor based on either the same or a different platform, offering significant flexibility in the design of the receiver. Our error analysis indicates that error rates below $0.3\%$ in the transduction step and gate errors below 0.2\% in the quantum processing will be required for first demonstrations of superadditivity, which is within reach of near-term hardware. 

We have established an efficient quantum receiver architecture and provided the first quantitative characterization of how realistic operational noise impacts superadditivity in quantum-assisted classical communications. Further research is needed to make quantum-assisted classical communications of practical use and our work points to several fruitful directions to be addressed. To extend this work, we note that the gate errors in the quantum receiver effectively create a~\enquote{super channel} on top of the pure loss bosonic channel originally considered for classical-quantum polar coding~\cite{saikat-wilde2}. The Holevo-Schumacher-Westmoreland (HSW) theorem implies that Holevo-bound achieving codes for this new noisy channel can be realized using a random-code constructed~\cite{hausladen, holevo} or using the general polar-coding technique applied to mixed-state modulation symbols~\cite{saikat-wilde2}, which would be interesting to pursue. 

In addition, our decoder in general requires non-Clifford gates. While work has been done on the efficacy of non-Clifford states and gates on the {\em quantum} capacity of particular quantum channels~\cite{bu}, the necessity and sufficiency of non-Clifford operations at the receiver for a classical-quantum channel remains, to our knowledge, unexplored. This line of inquiry builds on similar questions raised in terms of the role of non-Gaussian operations in the continuous variable setting, for all-photonic designs of Holevo-capacity-achieving receivers~\cite{takeoka}. 

Another unsolved problem of paramount practical importance is how the optical communications capacity, in the low mean photon flux regime (${\bar n} \ll 1$), scales with the code blocklength $N$ over which the quantum joint-detection receiver acts. In fact, even for the binary phase shift keying (BPSK) modulation alphabet of ${\bar n}$ mean received photon number per mode considered in this paper, we only know the capacity expressions for $C_1({\bar n}) = 1 - h_2([1-\sqrt{e^{-4{\bar n}}}]/2)$ and $C_\infty({\bar n}) = h_2([1-e^{-2{\bar n}}]/2)$, corresponding to $N=1$ (symbol by symbol receiver measurement, Dolinar receiver being optimum) and $N = \infty$ (Holevo limit), respectively. Quantifying the optimal increase of $C_N({\bar n})$ with $N$ (for relatively small values of $N$) and achieving $C_N$ with an appropriate code and a generalization of our proposed receiver design, is an important future research direction. 

Finally, we note that the photon-to-quantum-processor map that forms the core of our receiver design can be applied to various other use cases of photonic quantum information processing at the quantum limit, such as: squeezed-light-enhanced parameter estimation~\cite{grace}, object classification using a quantum-radar transmitter~\cite{cox}, quantum-assisted reading from a classical optical memory~\cite{wilde} and more.

\section*{Acknowledgments}
We thank Wastu Wisesa Ginanjar for assistance in the early stages of the project and Babak Saif and Nickholas Gutierrez for useful discussions. J.B. acknowledges support from The AWS Quantum Discovery Fund at the Harvard Quantum Initiative. S.G. acknowledges NASA for funding support. K.W.S. acknowledges support from the Natural Sciences and Engineering Research Council of Canada (NSERC) through a PGS D fellowship.

Distribution Statement A. Approved for public release. Distribution is unlimited.  This material is based upon work supported by the Under Secretary of Defense for Research and Engineering under Air Force Contract No. FA8702-15-D-0001 or FA8702-25-D-B002. Any opinions, findings, conclusions or recommendations expressed in this material are those of the author(s) and do not necessarily reflect the views of the Under Secretary of Defense for Research and Engineering. © 2025 Massachusetts Institute of Technology. Delivered the U.S. Government with Unlimited Rights, as defined in DFARS Part 252.227-7013 or 7014 (Feb 2014). Notwithstanding any copyright notice, U.S. Government rights in this work are defined by DFARS 252.227-7013 or DFARS 252.227-7014 as detailed above. Use of this work other than as specifically authorized by the U.S. Government may violate any copyrights that exist in this work.

\nocite{sorenson, arikanlecture}

\bibliography{sources}

\end{document}


\maketitle

\tableofcontents

\section{Projectors, probabilities and decoding, $N=4$}
To calculate the rate achieved by the CQ polar coded message with the quantum SC decoder, we calculate the mutual information between input and output and divide by the number of channel uses, then optimize over all possible input probability distributions. To calculate the mutual information, we need the positive-operator valued measure (POVM) for the SC decoder.

Let's find our POVM for $N=4, K=2$. Our information bits take on the values 00, 01, 10 or 11. We freeze the left-most two bits in the total input string $u = u_1 u_2 u_3 u_4$ to be 0 i.e. $u_1 = u_2 = 0$. So our input messages are: $ u \in \{0000, 0001, 0010, 0011\}$ with corresponding codewords: $\ket{\psi_{0000}}, \ket{\psi_{1111}}, \ket{\psi_{1010}}, \ket{\psi_{0101}}$. (The transformation to get the codewords $x$ from the input messages $u$ is $x_1 = u_1 \oplus u_2 \oplus u_3 \oplus u_4, x_2 = u_2 \oplus u_4, x_3 = u_3 \oplus u_4, x_4 = u_4$ \cite{arikan, arikanlecture}.)

To form the POVM for the successive cancellation decoder, note that our procedure is to first project onto $\Pi_{ijk}$, then onto $\Pi_{ijkl}$. We do a collective measurement over all modes to determine the left-most non-frozen bit. Then, we perform another collective measurement over all modes to determine the next non-frozen bit, assuming the result of the preceding measurement.

For example, if we want to find the probability that we detect outcome $y=0001$ given input message $u=0001$, we first compute: $\ket{\psi'} = \Pi_{000}\ket{\psi_{1111}}$, then \[P(y=0001|u=0001) = \bra{\psi'}\Pi_{0001}\ket{\psi'} = \bra{\psi_{1111}}\Pi_{000}\Pi_{0001}\Pi_{000}\ket{\psi_{1111}} = \bra{\psi_{1111}}\Lambda_{0001}\ket{\psi_{1111}}\]

These $\Lambda_{ijkl}$ represent the action of the SCD.

Following \cite{saikat-wilde1, saikat-wilde2}, our POVM is thus:

\[\Lambda_{0000} = \Pi_{000}\Pi_{0000}\Pi_{000}\]

\[\Lambda_{0001} = \Pi_{000}\Pi_{0001}\Pi_{000}\]

\[\Lambda_{0010} = \Pi_{001}\Pi_{0010}\Pi_{001}\]

\[\Lambda_{0011} = \Pi_{001}\Pi_{0011}\Pi_{001}\]

where the $\Pi_u$ are projectors defined as so:

\[\Pi_{(i)u_1^{i-1}0} = \{\overline{\rho}_{u_1^{i-1}0} - \overline{\rho}_{u_1^{i-1}1} \geq 0\} \text{ projection onto non-negative eigenspace}\]

\[\Pi_{(i)u_1^{i-1}1} = \{\overline{\rho}_{u_1^{i-1}0} - \overline{\rho}_{u_1^{i-1}1} < 0\} = I - \Pi_{(i)u_1^{i-1}0} \text{ projection onto negative eigenspace}\]

\[\overline{\rho}_{u_1^{k}} = \sum_{u_{k+1}^N}{\frac{1}{2^{N-k}} \ket{\psi_{u^NG^N}}\bra{\psi_{u^NG^N}}}\]

\[u^4G^4 = [u_1 \oplus u_2 \oplus u_3 \oplus u_4, u_2 \oplus u_4, u_3 \oplus u_4, u_4 ]\]

To get some intuition for why the projectors are formulated via an eigenspace projection onto the difference between density matrices for the $i^{th}$ bit equal to 0 and the $i^{th}$ bit equal to 1, think of trying to maximize the contrast between the state where the $i^{th}$ bit is 0 and the state where the $i^{th}$ bit is 1. In the 1 qubit limit where $\overline{\rho}_0$ and $\overline{\rho}_1$ are orthogonal pure states, i.e. $\overline{\rho}_0 = \left[\begin{matrix}1 & 0 \\0 & 0 \end{matrix}\right]$, $\overline{\rho}_1 = \left[\begin{matrix}0 & 0 \\0 & 1 \end{matrix}\right]$, this formulation gives:

\[\overline{\rho}_0 - \overline{\rho}_1 = \left[\begin{matrix}1 & 0 \\0 & 0 \end{matrix}\right] - \left[\begin{matrix}0 & 0 \\0 & 1 \end{matrix}\right] = \left[\begin{matrix}1 & 0 \\0 & -1 \end{matrix}\right]\]

which yields the expected Z projectors from the eigenspaces.

\subsection{Projectors for $i = 3$}

\[\overline{\rho}_{u_1^20} = \sum_{u_4 = 0}^1 \frac{1}{2}\ket{\psi_{u_1 \oplus u_2 \oplus 0 \oplus u_4 }\psi_{u_2 \oplus u_4}\psi_{0 \oplus u_4}\psi_{u_4}}\bra{\psi_{u_1 \oplus u_2 \oplus 0 \oplus u_4 }\psi_{u_2 \oplus u_4}\psi_{0 \oplus u_4}\psi_{u_4}} \]
\[= \frac{1}{2}\ket{\psi_{u_1 \oplus u_2 \oplus 0 \oplus 0 }\psi_{u_2 \oplus 0}\psi_{0 \oplus 0}\psi_{0}}\bra{\psi_{u_1 \oplus u_2 \oplus 0 \oplus 0 }\psi_{u_2 \oplus 0}\psi_{0 \oplus 0}\psi_{0}} + \frac{1}{2}\ket{\psi_{u_1 \oplus u_2 \oplus 0 \oplus 1 }\psi_{u_2 \oplus 1}\psi_{0 \oplus 1}\psi_{1}}\bra{\psi_{u_1 \oplus u_2 \oplus 0 \oplus 1 }\psi_{u_2 \oplus 1}\psi_{0 \oplus 1}\psi_{1}}\]

\[\overline{\rho}_{u_1^21} = \sum_{u_4 = 0}^1 \frac{1}{2}\ket{\psi_{u_1 \oplus u_2 \oplus 1 \oplus u_4 }\psi_{u_2 \oplus u_4}\psi_{1 \oplus u_4}\psi_{u_4}}\bra{\psi_{u_1 \oplus u_2 \oplus 1 \oplus u_4 }\psi_{u_2 \oplus u_4}\psi_{1 \oplus u_4}\psi_{u_4}} \]
\[= \frac{1}{2}\ket{\psi_{u_1 \oplus u_2 \oplus 1 \oplus 0 }\psi_{u_2 \oplus 0}\psi_{1 \oplus 0}\psi_{0}}\bra{\psi_{u_1 \oplus u_2 \oplus 1 \oplus 0 }\psi_{u_2 \oplus 0}\psi_{1 \oplus 0}\psi_{0}} + \frac{1}{2}\ket{\psi_{u_1 \oplus u_2 \oplus 1 \oplus 1 }\psi_{u_2 \oplus 1}\psi_{1 \oplus 1}\psi_{1}}\bra{\psi_{u_1 \oplus u_2 \oplus 1 \oplus 1 }\psi_{u_2 \oplus 1}\psi_{1 \oplus 1}\psi_{1}}\]

So for $\Pi_{000}$ and $\Pi_{001}$ we project onto the positive and negative eigenspaces respectively of:

\[\frac{1}{2}\left(\ket{\psi_0\psi_0\psi_0\psi_0}\bra{\psi_0\psi_0\psi_0\psi_0} + \ket{\psi_1\psi_1\psi_1\psi_1}\bra{\psi_1\psi_1\psi_1\psi_1} \right) - \frac{1}{2} \left(\ket{\psi_1\psi_0\psi_1\psi_0}\bra{\psi_1\psi_0\psi_1\psi_0} + \ket{\psi_0\psi_1\psi_0\psi_1}\bra{\psi_0\psi_1\psi_0\psi_1} \right) \]

If we condition on being in a 0-1 photon subspace of our Hilbert space (we  account for the effect of neglecting 2 or more photon events; see main text), this yields:

Eigenvectors (un-normalized):

Eigenvalue $= 0$:

\[\left[\begin{matrix}1 \\ 0 \\ 0 \\ 0 \\ 0\end{matrix}\right], \left[\begin{matrix}0 \\ -1 \\ 0 \\ 1 \\0\end{matrix}\right], \left[\begin{matrix}0 \\ 0 \\ -1 \\ 0 \\1 \end{matrix}\right]\]

Eigenvalue $> 0$:

\[\left[\begin{matrix}0 \\ 1 \\ 1 \\ 1 \\ 1\end{matrix}\right]\]

Eigenvalue $< 0$:

\[\left[\begin{matrix}0 \\ -1 \\ 1 \\ -1 \\ 1\end{matrix}\right]\]

Projection onto eigenspaces:

\[ \Pi_{000} = 
\left[\begin{matrix}1 & 0 & 0 & 0 & 0\\0 & 0.75 & 0.25 & -0.25 & 0.25\\0 & 0.25 & 0.75 & 0.25 & -0.25\\0 & -0.25 & 0.25 & 0.75 & 0.25\\0 & 0.25 & -0.25 & 0.25 & 0.75\end{matrix}\right]
, \Pi_{001} = 
\left[\begin{matrix}0 & 0 & 0 & 0 & 0\\0 & 0.25 & -0.25 & 0.25 & -0.25\\0 & -0.25 & 0.25 & -0.25 & 0.25\\0 & 0.25 & -0.25 & 0.25 & -0.25\\0 & -0.25 & 0.25 & -0.25 & 0.25\end{matrix}\right]
\]

Where we have a 5D (truncated) subspace with, for example:

\[\ket{\psi_{0000}} = \frac{1}{1+4\alpha^2}[1, \alpha, \alpha, \alpha, \alpha]^T = \frac{1}{1+ 4\alpha^2} \left[ \ket{0000} + \alpha \left(\ket{0001} + \ket{0010} + \ket{0100} + \ket{1000} \right) \right]\]

\subsection{Projectors for $i = 4$}

\[\overline{\rho}_{u_1^30} = \ket{\psi_{u_1 \oplus u_2 \oplus u_3 \oplus 0 }\psi_{u_2 \oplus 0}\psi_{u_3 \oplus 0}\psi_{0}}\bra{\psi_{u_1 \oplus u_2 \oplus u_3 \oplus 0 }\psi_{u_2 \oplus 0}\psi_{u_3 \oplus 0}\psi_{0}}\]

\[\overline{\rho}_{u_1^31} = \ket{\psi_{u_1 \oplus u_2 \oplus u_3 \oplus 1 }\psi_{u_2 \oplus 1}\psi_{u_2 \oplus 1}\psi_{1}}\bra{\psi_{u_1 \oplus u_2 \oplus u_3 \oplus 1 }\psi_{u_2 \oplus 1}\psi_{u_3 \oplus 0}\psi_{1}}\]

For $\Pi_{0000}$ and $\Pi_{0001}$ we project onto the positive and negative eigenspaces respectively of:

\[\ket{\psi_0\psi_0\psi_0\psi_0}\bra{\psi_0\psi_0\psi_0\psi_0} - \ket{\psi_1\psi_1\psi_1\psi_1}\bra{\psi_1\psi_1\psi_1\psi_1} \]

and for $\Pi_{0010}$ and $\Pi_{0011}$ we project onto the positive and negative eigenspaces respectively of:

\[\ket{\psi_1\psi_0\psi_1\psi_0}\bra{\psi_1\psi_0\psi_1\psi_0} - \ket{\psi_0\psi_1\psi_0\psi_1}\bra{\psi_0\psi_1\psi_0\psi_1} \]

If we condition on being in a 0-1 photon subspace of our Hilbert space, this yields:

\[ \Pi_{0000} = 
\left[\begin{matrix}\frac{1}{2} & \frac{1}{4} & \frac{1}{4} & \frac{1}{4} & \frac{1}{4}\\\frac{1}{4} & \frac{7}{8} & - \frac{1}{8} & - \frac{1}{8} & - \frac{1}{8}\\\frac{1}{4} & - \frac{1}{8} & \frac{7}{8} & - \frac{1}{8} & - \frac{1}{8}\\\frac{1}{4} & - \frac{1}{8} & - \frac{1}{8} & \frac{7}{8} & - \frac{1}{8}\\\frac{1}{4} & - \frac{1}{8} & - \frac{1}{8} & - \frac{1}{8} & \frac{7}{8}\end{matrix}\right]
, \Pi_{0001} = 
\left[\begin{matrix}\frac{1}{2} & - \frac{1}{4} & - \frac{1}{4} & - \frac{1}{4} & - \frac{1}{4}\\- \frac{1}{4} & \frac{1}{8} & \frac{1}{8} & \frac{1}{8} & \frac{1}{8}\\- \frac{1}{4} & \frac{1}{8} & \frac{1}{8} & \frac{1}{8} & \frac{1}{8}\\- \frac{1}{4} & \frac{1}{8} & \frac{1}{8} & \frac{1}{8} & \frac{1}{8}\\- \frac{1}{4} & \frac{1}{8} & \frac{1}{8} & \frac{1}{8} & \frac{1}{8}\end{matrix}\right]
\]

\[ \Pi_{0010} = 
\left[\begin{matrix}\frac{1}{2} & - \frac{1}{4} & \frac{1}{4} & - \frac{1}{4} & \frac{1}{4}\\- \frac{1}{4} & \frac{7}{8} & \frac{1}{8} & - \frac{1}{8} & \frac{1}{8}\\\frac{1}{4} & \frac{1}{8} & \frac{7}{8} & \frac{1}{8} & - \frac{1}{8}\\- \frac{1}{4} & - \frac{1}{8} & \frac{1}{8} & \frac{7}{8} & \frac{1}{8}\\\frac{1}{4} & \frac{1}{8} & - \frac{1}{8} & \frac{1}{8} & \frac{7}{8}\end{matrix}\right]
, \Pi_{0011} = 
\left[\begin{matrix}\frac{1}{2} & \frac{1}{4} & - \frac{1}{4} & \frac{1}{4} & - \frac{1}{4}\\\frac{1}{4} & \frac{1}{8} & - \frac{1}{8} & \frac{1}{8} & - \frac{1}{8}\\- \frac{1}{4} & - \frac{1}{8} & \frac{1}{8} & - \frac{1}{8} & \frac{1}{8}\\\frac{1}{4} & \frac{1}{8} & - \frac{1}{8} & \frac{1}{8} & - \frac{1}{8}\\- \frac{1}{4} & - \frac{1}{8} & \frac{1}{8} & - \frac{1}{8} & \frac{1}{8}\end{matrix}\right]
\]

\subsection{POVM for $N=4, K=2$} 

Putting the results for the projectors together we find:

\[\Lambda_{0000} = \Pi_{000}\Pi_{0000}\Pi_{000} =
\left[\begin{matrix}\frac{1}{2} & 0.25 & 0.25 & 0.25 & 0.25\\0.25 & 0.625 & 0.125 & -0.375 & 0.125\\0.25 & 0.125 & 0.625 & 0.125 & -0.375\\0.25 & -0.375 & 0.125 & 0.625 & 0.125\\0.25 & 0.125 & -0.375 & 0.125 & 0.625\end{matrix}\right]
\]

\[\Lambda_{0001} = \Pi_{000}\Pi_{0001}\Pi_{000} = 
\left[\begin{matrix}\frac{1}{2} & -0.25 & -0.25 & -0.25 & -0.25\\-0.25 & 0.125 & 0.125 & 0.125 & 0.125\\-0.25 & 0.125 & 0.125 & 0.125 & 0.125\\-0.25 & 0.125 & 0.125 & 0.125 & 0.125\\-0.25 & 0.125 & 0.125 & 0.125 & 0.125\end{matrix}\right]
\]

\[\Lambda_{0010} = \Pi_{001}\Pi_{0010}\Pi_{001} = 
\left[\begin{matrix}0 & 0 & 0 & 0 & 0\\0 & 0.125 & -0.125 & 0.125 & -0.125\\0 & -0.125 & 0.125 & -0.125 & 0.125\\0 & 0.125 & -0.125 & 0.125 & -0.125\\0 & -0.125 & 0.125 & -0.125 & 0.125\end{matrix}\right]
\]

\[\Lambda_{0011} = \Pi_{001}\Pi_{0011}\Pi_{001} = 
\left[\begin{matrix}0 & 0 & 0 & 0 & 0\\0 & 0.125 & -0.125 & 0.125 & -0.125\\0 & -0.125 & 0.125 & -0.125 & 0.125\\0 & 0.125 & -0.125 & 0.125 & -0.125\\0 & -0.125 & 0.125 & -0.125 & 0.125\end{matrix}\right]
\]

One can verify that the sum over all elements of the POVM is $I$.

Note that 0010 and 0011 have identical projectors so these messages cannot be distinguished. This will be reflected in the optimal input distribution as well.

\subsection{Transition table from the POVM}
We next calculate for each $\ket{\psi_{u^NG^N}}$

\[H_{\Lambda}(\ket{\psi_{u^NG^N}}) = -\sum_y{\bra{\psi_{u^NG^N}}\Lambda_y\ket{\psi_{u^NG^N}}}\log\bra{\psi_{u^NG^N}}\Lambda_y\ket{\psi_{u^NG^N}}\]

\noindent where $\bra{\psi_{u^NG^N}}\Lambda_y\ket{\psi_{u^NG^N}}$ is the probability of measuring outcome $y$ if in the state $\ket{\psi_{u^NG^N}}$ and $H_{\Lambda}(\ket{\psi_{u^NG^N}})$ is the entropy of the distribution induced by the POVM $\Lambda = \{ \Lambda_y\}$ conditioned on sending the state $\ket{\psi_{u^NG^N}}$. The distribution for each $\ket{\psi_{u^NG^N}}$ is given in Table 1.

We have again assumed a truncated 0-1 photon subspace.

\bigskip

\begin{table}[H]
 \centering
 \caption{The distribution induced by the POVM $\Lambda = \{ \Lambda_y\}$ for each state $\ket{\psi_{u^NG^N}}$.}
\begin{tabular}{ |p{3cm}|p{1.5cm}|p{1.5cm}|p{1.5cm}|p{1.5cm}|}
\hline
 $\bra{\psi_{u^NG^N}}\Lambda_y\ket{\psi_{u^NG^N}}$& $\Lambda_{0000}$ & $\Lambda_{0001}$ &$\Lambda_{0010}$ & $\Lambda_{0011}$ \\
 \hline
 $\ket{\psi_{0000}}, u = 0000$ & $\frac{2\alpha^2 + 2\alpha + 0.5}{4\alpha^2 + 1}$ &  $\frac{2\alpha^2 - 2\alpha + 0.5}{4\alpha^2 + 1}$  &   0 & 0\\
 $\ket{\psi_{1111}}, u=0001$ & $\frac{2\alpha^2 - 2\alpha + 0.5}{4\alpha^2 + 1}$  &  $\frac{2\alpha^2 + 2\alpha + 0.5}{4\alpha^2 + 1}$  &   0 & 0\\
$\ket{\psi_{1010}}, u=0010$ & $\frac{0.5}{4\alpha^2+1}$  &  $\frac{0.5}{4\alpha^2+1}$  &   $\frac{2 \alpha^2}{4\alpha^2+1}$ & $\frac{2 \alpha^2}{4\alpha^2+1}$\\
 $\ket{\psi_{0101}}, u=0011$ & $\frac{0.5}{4\alpha^2+1}$  &  $\frac{0.5}{4\alpha^2+1}$  &   $\frac{2 \alpha^2}{4\alpha^2+1}$ & $\frac{2 \alpha^2}{4\alpha^2+1}$\\
 \hline
\end{tabular}
\end{table}

\subsection{Optimizing the input distribution}
Now that we have the transition table (Table 1) what remains is to find the input distribution that maximizes our rate. Our unknowns are the 4 probabilities $p(u=0000), p(u=0001), p(u=0010), p(u=0011)$, subject to non-negativity and sum to unity constraints.

First, we note that the projectors and therefore distributions induced by $\Lambda_{0010}$ and $\Lambda_{0011}$ are identical (see the previous section). Due to this, the expression for mutual information will depend on the \emph{sum} of $p(u=0010), p(u=0011)$ and not on each value individually, as we show in the following paragraphs.

\medskip

We can rewrite the transition table in the previous section in the following way, with $q =$ and $k =$:

\begin{table}[H]
\centering
\caption{The distribution induced by the POVM $\Lambda = \{ \Lambda_y\}$ for each state $\ket{\psi_{u^NG^N}}$ in a simpler form.}
\begin{tabular}{ |p{3cm}|p{1.5cm}|p{1.5cm}|p{1.5cm}|p{1.5cm}|}
 \hline
 $P(y|u)$ & $y = 0000$ & $y = 0001 $ &$y = 0010$ & $y = 0011 $ \\
 \hline
 $u = 0000$ & $q$ &  $1-q$  &   0 & 0\\
 $u=0001$ & $1-q$  &  $q$  &   0 & 0\\
 $u=0010$ & $k$  &  $k$  &   $\frac{1}{2} - k$ & $\frac{1}{2} - k$\\
 $u=0011$ & $k$  &  $k$  &   $\frac{1}{2} - k$ & $\frac{1}{2} - k$\\
 \hline
\end{tabular}
\end{table}

\medskip

Our expression for mutual information is

\[I(Y;U) = H(Y) - H(Y|U) \quad (\star)\]

Let $p_u = [p_0, p_1, p_2, p_3]$.

First, we calculate the distribution for the output message $Y$:

\begin{gather*}
    P(Y = y) = \sum_u{p_U(u)p_{Y|U}(Y = y|U = u) } \\
    P(Y = 0000) = p_0 q + p_1(1-q) + k p_2 + k p_3 \\
    P(Y = 0001) = p_0 (1-q) + p_1 q + k p_2 + k p_3 \\
    P(Y = 0010) = P(Y = 0011) = (\frac{1}{2} -k)(p_2+p_3)  \\
\end{gather*}

Then, we calculate the entropy of $Y$:

\begin{align*}
    H(Y) =& - \sum_y{p_y \log{p_y}}\\
    =& -[p_0q + p_1(1-q) + k(p_2+p_3)]\log[p_0q + p_1(1-q) + k(p_2+p_3)] \\
    &- [p_0(1-q) + p_1 q + k(p_2+p_3)]\log[p_0(1-q) + p_1 q + k(p_2+p_3)] - 2[(\frac{1}{2}-k)(p_2+p_3)]\log{[(\frac{1}{2}-k)(p_2+p_3)]}\\
\end{align*}

\noindent and the entropy of $Y$ given $U$:

\begin{gather*}
    H(Y|U) = -\sum_u{p_u H(Y|U=u)}\\
    H(Y|U=u) = -\sum_y{p(y|U=u)\log{p(y|U=u)}}\\
    H(Y|U=0000) = H(Y|U=0001) = -q\log{q} -(1-q)\log{(1-q)}\\
    H(Y|U=0010) = H(Y|U=0011) = -2k\log k - 2(\frac{1}{2}-k)\log{(\frac{1}{2}-k)}
\end{gather*}

We sub these expressions into $(\star)$:

\begin{align*}
    I(Y;U) = -[p_0q + p_1(1-q) + k(p_2+p_3)]\log[p_0q + p_1(1-q) + k(p_2+p_3)] - [p_0(1-q) + p_1 q + \\
    k(p_2+p_3)]\log[p_0(1-q) + p_1 q + k(p_2+p_3)] - 2[(\frac{1}{2}-k)(p_2+p_3)]\log{[(\frac{1}{2}-k)(p_2+p_3)]}\\
    + (p_0+p_1)[q\log{q} +(1-q)\log{(1-q}] \\
    + (p_2+p_3)[2k\log{k} + 2(\frac{1}{2}-k)\log{(\frac{1}{2}-k)}]
\end{align*}

The expression for mutual information depends on the sum of $p_2$ and $p_3$ and not on each value individually. As such, the optimal distribution is a set for which the sum $p_2+p_3$ is constant.

Next, we want to find the $p_u$ that minimize $(\star)$. We perform the optimization numerically using Python's scipy package and find that the optimal distribution is: $\{0.48, 0.48, p_2, p_3\}$ such that the sum of $p_2, p_3$ is 0.04. We pick $\{0.48, 0.48, 0.04, 0 \}$ since 0010 and 0011 cannot be distinguished.

In this case, without additional errors, we optimized with the mean photon number fixed at 0.001, since the curves are monotonic. As detailed in the main text, we also add errors to the receiving process. In this case, we will perform the optimization for each average photon number (or for each $\alpha$) since the behavior is not necessarily monotonic.

\subsection{Determination of the decoding circuits}

To determine the gates required to implement the quantum SC projectors, we find a set of unitary gates that transform the memory state such that a Z basis measurement decodes the $i^{th}$ bit as 0 or 1. For example, for $N=4$, we take one of the 4th bit projectors:
    
\begin{align*}
    \Pi_{0001} =& 
    \left[\begin{matrix}\frac{1}{2} & - \frac{1}{4} & - \frac{1}{4} & - \frac{1}{4} & - \frac{1}{4}\\- \frac{1}{4} & \frac{1}{8} & \frac{1}{8} & \frac{1}{8} & \frac{1}{8}\\- \frac{1}{4} & \frac{1}{8} & \frac{1}{8} & \frac{1}{8} & \frac{1}{8}\\- \frac{1}{4} & \frac{1}{8} & \frac{1}{8} & \frac{1}{8} & \frac{1}{8}\\- \frac{1}{4} & \frac{1}{8} & \frac{1}{8} & \frac{1}{8} & \frac{1}{8}\end{matrix}\right]\\
    =& \biggl[-\frac{1}{\sqrt{2}}\ket{vac} + \frac{1}{2 \sqrt{2}}\bigl[\ket{1} + \ket{2} + \ket{3} + \ket{4}\bigr]\biggr]\biggl[ -\frac{1}{\sqrt{2}}\bra{vac} + \frac{1}{2 \sqrt{2}}\bigl[\bra{1} + \bra{2} + \bra{3} + \bra{4} \bigr]\biggr] \\
    =& \ket{\psi_{\Pi_{0001}}}\bra{\psi_{\Pi_{0001}}}\\[4em]
    \Pi_{001} =&
    \left[\begin{matrix}0 & 0 & 0 & 0 & 0\\0 & 0.25 & -0.25 & 0.25 & -0.25\\0 & -0.25 & 0.25 & -0.25 & 0.25\\0 & 0.25 & -0.25 & 0.25 & -0.25\\0 & -0.25 & 0.25 & -0.25 & 0.25\end{matrix}\right]\\
    =&\frac{1}{4}\biggl[\ket{1} - \ket{2} + \ket{3} - \ket{4}\biggr]\biggl[\bra{1} - \bra{2} + \bra{3} - \bra{4}\biggr] \\
    =&  \ket{\psi_{\Pi_{001}}}\bra{\psi_{\Pi_{001}}}
\end{align*}

\noindent We then set a time bin encoding:

\begin{align*}
    \ket{\uparrow \uparrow \uparrow} = \ket{vac}\\
    \ket{\downarrow \uparrow \uparrow} = \ket{1}\\
    \ket{\downarrow \uparrow \downarrow} = \ket{2}\\
    \ket{\downarrow \downarrow \uparrow} = \ket{3}\\
    \ket{\downarrow \downarrow \downarrow} = \ket{4}\\
\end{align*}

\noindent In general, the time bin encoding is arbitrary and should be determined based on what makes decoding most simple for the specific physical implementation used.

Once the time bin encoding is set, the next step is to determine gates that take the superposition to a single computational basis state. We first rearrange the expression for $\ket{\psi_{\Pi_{0001}}}$ to make the action of the gates more apparent:

\begin{align*}
    \ket{\psi_{\Pi_{0001}}} =& -\frac{1}{\sqrt{2}}\ket{\uparrow \uparrow \uparrow} + \frac{1}{2 \sqrt{2}}\bigl[\ket{\downarrow \uparrow \uparrow} + \ket{\downarrow \uparrow \downarrow} + \ket{\downarrow \downarrow \uparrow} + \ket{\downarrow \downarrow \downarrow}\bigr] \\
    =& -\frac{1}{\sqrt{2}}\ket{\uparrow \uparrow \uparrow} + \frac{1}{2 \sqrt{2}} \ket{\downarrow} \otimes \bigl[\ket{ \uparrow \uparrow} + \ket{ \uparrow \downarrow} + \ket{ \downarrow \uparrow} + \ket{\downarrow \downarrow}\bigr] \\
    =& -\frac{1}{\sqrt{2}}\ket{\uparrow \uparrow \uparrow} + \frac{1}{2 \sqrt{2}} \ket{\downarrow} \otimes \bigl[(\ket{ \uparrow } + \ket{ \downarrow}) \otimes \ket{\uparrow} + (\ket{ \uparrow } + \ket{\downarrow}) \otimes \ket{\downarrow}\bigr] 
\end{align*}

\noindent A controlled Hadamard with qubit 0 as the control and qubit 1 as the target gives:

\[-\frac{1}{\sqrt{2}}\ket{\uparrow \uparrow \uparrow} + \frac{1}{2}\bigl[\ket{\downarrow \uparrow \uparrow} + \ket{\downarrow \uparrow \downarrow} \bigr] = -\frac{1}{\sqrt{2}}\ket{\uparrow \uparrow \uparrow} + \frac{1}{2} \ket{\downarrow \uparrow} \otimes \bigl[\ket{\uparrow} + \ket{\downarrow } \bigr]\]

\noindent Next, a controlled Hadamard with qubit 0 as the control and qubit 2 as the target yields:

\[-\frac{1}{\sqrt{2}}\biggl[\ket{\uparrow \uparrow \uparrow} - \ket{\downarrow \uparrow \uparrow} \biggr]\]

Finally, a $H$ gate on qubit 0 gives:

\[-\ket{\downarrow \uparrow \uparrow}\]

To determine the qubits to measure, we first find the bit 4 projector circuit in the same way as above, then input the codewords to see which bits help us distinguish the states.

\subsection{Intuition behind the decoding circuit}
Analyzing the possible codeword states provides intuition for how the decoding circuit works. They can be grouped into pairs, where the codewords within a pair differ only in the relative phase between the vacuum and the presence of a photon. For example, for $N=4, K=2$, there are two sets of codeword pairs:

\begin{align*}
\ket{\psi_{1111}} \approx \ket{vac} -\alpha \left(  \ket{1} +  \ket{2}+ \ket{3} +  \ket{4}\right)\\
\ket{\psi_{0000}} \approx \ket{vac} +\alpha \left(  \ket{1} +  \ket{2}+ \ket{3} +  \ket{4}\right)
\end{align*}

\noindent and

\begin{align*}
\ket{\psi_{1010}} \approx \ket{vac} - \alpha \left(\ket{1} - \ket{2} + \ket{3} - \ket{4} \right) \\
\ket{\psi_{0101}} \approx \ket{vac} + \alpha \left(\ket{1} - \ket{2} + \ket{3} - \ket{4} \right)
\end{align*}

The decoding circuit repetitively bifurcates the codespace based on the relative phases between the state components, until the last step is to decide on a state within the final pair.

These symmetric states also lead to symmetry in the optimal input distributions.

\section{Decoding circuit and procedure for $N=8$}

To find the projectors and decoding procedure for $N=8$, we proceed similarly to $N=4$, following Refs. \cite{saikat-wilde1} and \cite{saikat-wilde2}. In Figure 1 and Table 1, we detail the decision tree and decoding table for $N=8$.

\begin{figure}[H]
    \centering
    \includegraphics[height=8cm, width=15cm]{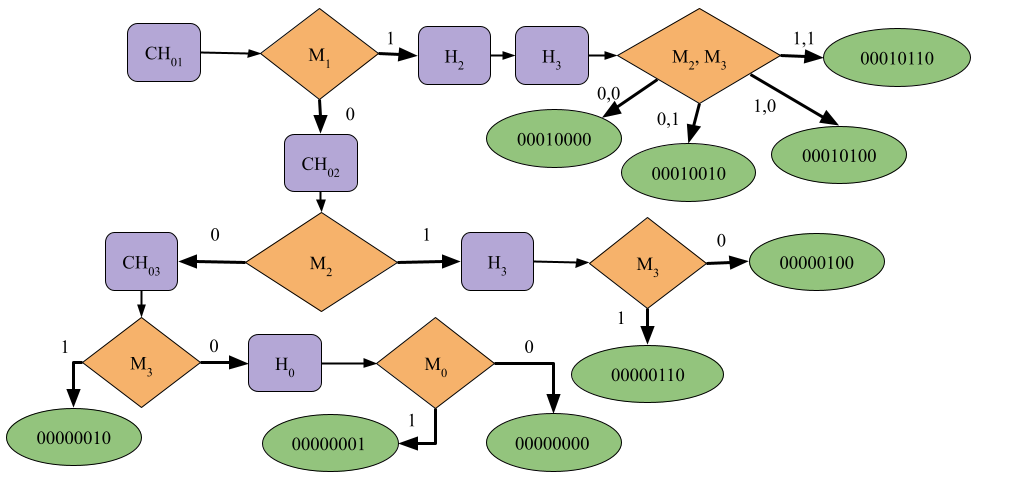}
    \caption{Decision tree for $N=8$. $M_i$ represents a measurement of qubit $i$ in the $Z$ basis. The binary labels denote measurement outcomes.$CH_{ij}$ is a controlled Hadamard gate with qubit $i$ as control and $j$ as target. $H_i$ is a Hadamard gate on qubit $i$. }
\end{figure}

\begin{table}[H]
    \centering
    \caption{Decoding table for $N=8, K=4$. The optimal distribution only has 9 codewords with nonzero input probability. The 4 bit binary strings not shown in this table would herald an error if measured.}
    \begin{tabular}{ |p{1.2cm}|p{1.2cm}|p{1.2cm}|p{1.2cm}|p{1.2cm}|}
     \hline
     Qubit 0 & Qubit 1 & Qubit 2 & Qubit 3 & Decision  \\
     \hline
     0 & 0  &  0 & 0 &   00000000 \\
     1 & 0  & 0  & 0 &   00000001  \\
     0 & 0  &  0 & 1 & 00000010 \\
     1 & 0 &  1 & 0 &  00000100 \\
     0 & 0  &  1 & 1 & 00000110 \\
     1 & 1 &  0 &  0 &   00010000 \\
     1 &  1 & 0  &  1 &  00010010 \\
     1 & 1  &  1 & 0 &   00010100 \\
     1 & 1  &  1 & 1 &   00010110 \\
     \hline
    \end{tabular}
\end{table}

The decoding procedure for $N=8$ involves more gates and operations than for $N=4$ due to the larger codespace. For ease of understanding, in this section we separate the graphical representation of the $N=8$ decoding circuit into four parts, each corresponding to a different set of paths through the decision tree in Figure 1. In the circuit diagrams, the topmost wire is qubit 0 and the bottommost is qubit 3.

\bigskip

\noindent Path 1:

\bigskip
\Qcircuit @C=1em @R=2em {
&  \qw  & \ctrl{1} & \qw & \qw \\
&  \qw & \gate{H} & \qw & \meter & \ustick{1} \\
&   \qw & \qw & \qw  & \qw & \qw & \gate{H} & \meter \\
&   \qw & \qw & \qw  & \qw & \qw & \gate{H} & \meter
}
\bigskip

\noindent Path 2:

\bigskip
\Qcircuit @C=1em @R=2em {
&  \qw  & \ctrl{1} & \qw & \qw & \qw & \ctrl{2} \\
&  \qw & \gate{H} & \qw & \meter & \ustick{0} \\
&   \qw & \qw & \qw  & \qw & \qw & \gate{H} & \meter & \ustick{1} \\
&   \qw & \qw & \qw  & \qw & \qw & \qw & \qw & \gate{H} & \meter 
}
\bigskip

\noindent Path 3:

\bigskip
\Qcircuit @C=1em @R=2em {
&  \qw  & \ctrl{1} & \qw & \qw & \qw & \ctrl{2} & \qw & \qw & \ctrl{3}   & \qw \\
&  \qw & \gate{H} & \qw & \meter & \ustick{0} \\
&   \qw & \qw & \qw  & \qw & \qw & \gate{H} & \meter & \ustick{0} \\
&   \qw & \qw & \qw  & \qw & \qw & \qw & \qw & \qw & \gate{H} & \meter & \ustick{1}  
}
\bigskip

\noindent Path 4:

\bigskip
\Qcircuit @C=1em @R=2em {
&  \qw  & \ctrl{1} & \qw & \qw & \qw & \ctrl{2} & \qw & \qw & \ctrl{3}   & \qw & \gate{H} & \meter\\
&  \qw & \gate{H} & \qw & \meter & \ustick{0} \\
&   \qw & \qw & \qw  & \qw & \qw & \gate{H} & \meter & \ustick{0} \\
&   \qw & \qw & \qw  & \qw & \qw & \qw & \qw & \qw & \gate{H} & \meter & \ustick{0}  
}
\bigskip

For $N=8$, the optimal distribution was found to be $\{0.325, 0.325, p_{11}, p_{12}, p_{21}, p_{22}, p_{31}, p_{32}, ..., p_{71}, p_{72} \}$ such that the sum of $p_{i1}$ and $p_{i2} = 0.05$

We pick: $\{0.325, 0.325, 0.05, 0, 0.05, 0, 0.05, 0, 0.05, 0, 0.05, 0, 0.05, 0, 0.05, 0\}$

\section{Storing more than one photon}

In the main text, we gave a high-level description of the multi-photon receiver. Here, we motivate the need for a receiver capable of storing multi-photon states and detail a possible implementation of the photon switch with a double-sided atom-cavity system.

To get a sense of the photon number of the states input to the receiver, we will look at two arbitrarily chosen code performances. We choose a rate $R = 1/2$ and a rate $R = 1/8$ code. We pick a source/information bit number of $K=4$ (not close to the Holevo bound) that is realistic for near-term experiments given the quantum hardware/resources required.

\begin{figure}[H]
    \centering
    \includegraphics[height=6cm, width=13cm]{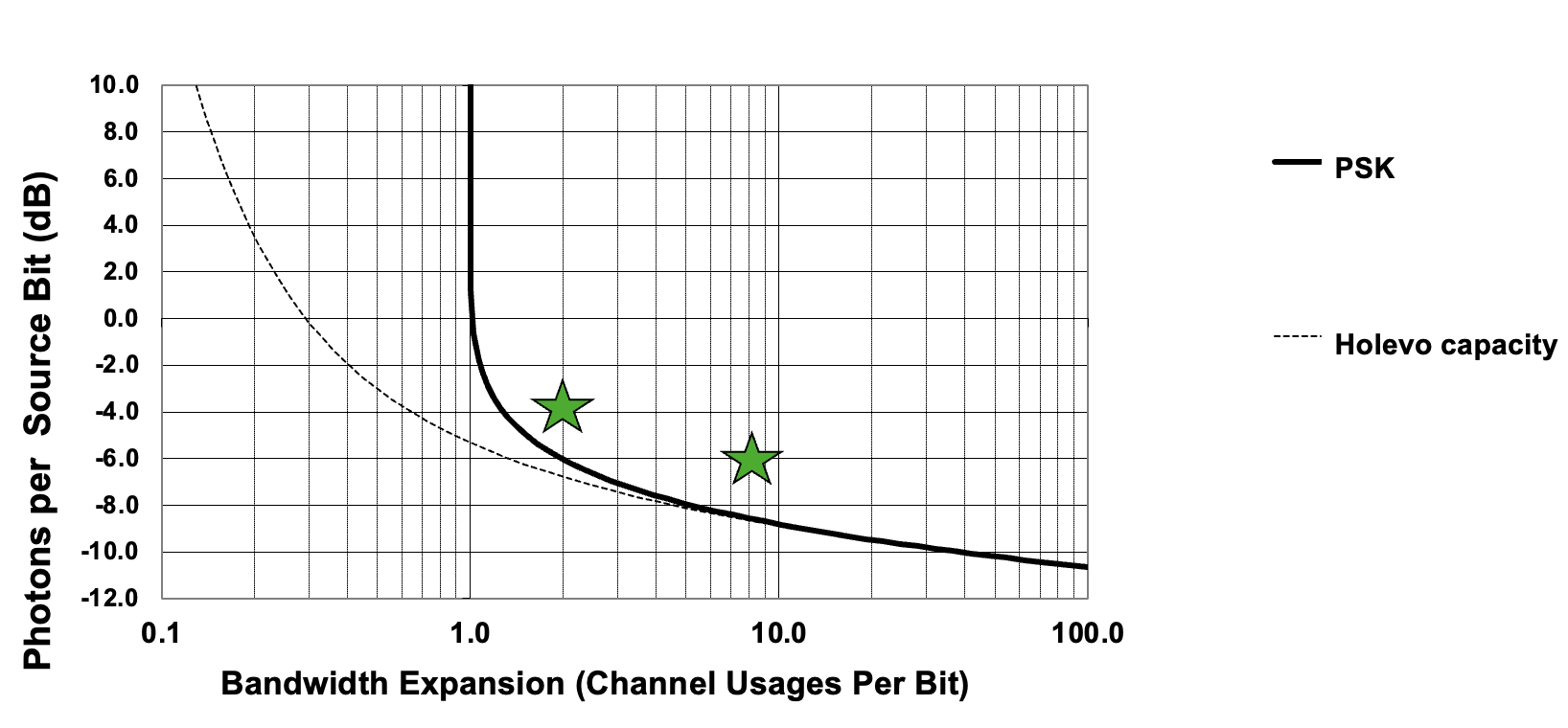}
    \caption{The stars mark possible performances for a rate $R = 1/2$ (left star) and a rate $R = 1/8$ code (right star).}
\end{figure}

For the $R=1/2$ code, we pick a performance of -4 dB photons per source bit. For the $R=1/8$ code, we pick a performance of -6 dB photons per source bit (Figure 2). These values are not at the Holevo capacity optimum, as we would expect for a small codeword size and with system imperfections. The number of photons per codeword for each performance is then:

\medskip

\noindent $R=1/2$, $\sim 0.4$ photons per source bit: At $K = 4$, 1.6 photons per codeword on average

\noindent $R=1/8$, $\sim 0.25$ photons per source bit: At $K = 4$, 1 photon per codeword on average

\medskip

Note that assuming Poisson statistics for the photon number, an average of 1 photon per codeword will have greater than 1 photon about one-third of the time. So, even for small blocklengths of $N=8$ and $N=32$ (implied by $K=4, R=1/2, R=1/8$), more than one photon might be received on average. 

\subsection{Quantum switch with a double-sided atom-cavity system}

\begin{figure}[H]
    \centering
    \includegraphics[height=4.5cm, width=6cm]{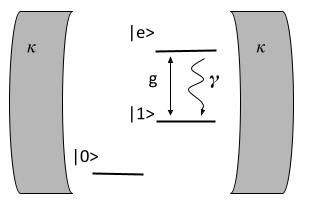}
    \caption{The double-sided atom cavity system.}
\end{figure}

A single atom with levels $\ket{0}, \ket{1}, \ket{e}$ is contained in a double-sided cavity with equal mirror transmittance $\kappa$ (Figure 3). The level $\ket{1}$ is coupled to $\ket{e}$ with coupling constant $g$ and an atom in $\ket{e}$ decays to $\ket{1}$ at rate $\gamma$. In the limit of at most a single photon present in the cavity at a time, and with the photon frequency resonant with the cavity mode and the cavity mode resonant with the atomic transition frequency, the probabilities of reflecting an incident photon in the coupled (atom in $\ket{1}$) and uncoupled (atom in $\ket{0}$) states are given by \cite{sorenson}:

\[R_{c} = \biggl(\frac{4C}{1+4C} \biggr)^2\]

\[R_{u} = 0\]

\noindent $C = \frac{g^2}{\kappa \gamma}$ is the cooperativity of the atom-cavity system. $C$ quantifies the ratio of the atom-photon interaction strength and the dissipative mechanisms (spontaneous emission and cavity loss).

If the atom is in $\ket{0}$, then $R = 0, T = 1$. If the atom is in $\ket{1}$, then $R$ is very large and $T$ is very small for large $C$. The system thus acts as a quantum switch. If the atom is in $\ket{0}$, the photon is transmitted through the cavity, but if the atom is in $\ket{1}$, it is reflected. This switch is used to selectively route photons to single photon receiver modules as described in the main text.

\subsection{Projectors for the multi-photon receiver}
The procedure for determining the projectors for the multi-photon receiver is very similar to that for the single-photon receiver, but with a larger Hilbert space. For example, for $N=4$ and 2 photons, we now have 6 qubits across both single-photon modules. Our 2 photon received state is of the form (taking the approximation up to 2 photons in different time bins):

\begin{align*}
\ket{\psi_{c}} =& \bigotimes_{i=1}^4\ket{(-1)^{c_i}\alpha} \\
\approx& \ket{0000} + \alpha [  (-1)^{c_1}\ket{1000} +  (-1)^{c_2}\ket{0100} + (-1)^{c_3}\ket{0010} +  (-1)^{c_4}\ket{0001} + (-1)^{c_1+c_2} \ket{1100}\\
&+ (-1)^{c_1+c_3} \ket{1010} + (-1)^{c_1+c_4} \ket{1001} + (-1)^{c_2+c_3} \ket{0110}+ (-1)^{c_2+c_4} \ket{0101} + (-1)^{c_3+c_4} \ket{0011}]
\end{align*}

To map to the memories, we can define the following time bin encoding:

\begin{align*}
    \ket{\uparrow \uparrow \uparrow, \uparrow \uparrow \uparrow} = \ket{0000}\\
    \ket{\downarrow \uparrow \uparrow, \uparrow \uparrow \uparrow} = \ket{1000}\\
    \ket{\downarrow \uparrow \downarrow, \uparrow \uparrow \uparrow} = \ket{0100}\\
    \ket{\downarrow \downarrow \uparrow, \uparrow \uparrow \uparrow} = \ket{0010}\\
    \ket{\downarrow \downarrow \downarrow, \uparrow \uparrow \uparrow} = \ket{0001}\\
    \ket{\downarrow \uparrow \uparrow, \downarrow \uparrow \downarrow} = \ket{1100}\\
    \ket{\downarrow \uparrow \uparrow, \downarrow \downarrow \uparrow} = \ket{1010}\\
    \ket{\downarrow \uparrow \uparrow, \downarrow \downarrow \downarrow} = \ket{1001}\\
    \ket{\downarrow \uparrow \downarrow, \downarrow \downarrow \uparrow} = \ket{0110}\\
    \ket{\downarrow \uparrow \downarrow, \downarrow \downarrow \downarrow} = \ket{0101}\\
    \ket{\downarrow \downarrow \uparrow, \downarrow \downarrow \downarrow} = \ket{0011}\\
\end{align*}

The idea here is that the first group of spins corresponds to the arrival time of the first photon, and the second group corresponds to the arrival time of the second photon.

We can then find the projectors in the same way as before (see Section 1 of this supplementary material), but using the new 2 photon codewords and encodings.

\section{Error modeling in detail}

To model errors, we consider a pure state $\rho = \ket{\psi}\bra{\psi}$ input to our receiver, with $\ket{\psi} \in \{\text{polar codewords} \}$. We then performed a series of operations on $\rho$ corresponding to each step in the receiver protocol. For operations with errors, we represented the transformations via the operator-sum or Kraus representation of a quantum channel, $\mathcal{K}$. In this representation, an input state $\rho$ is mapped to an output state $\rho'$ as follows:

\[\rho' = \mathcal{K}(\rho) = \sum_j A_j \rho A_j^{\dag}\]

\noindent where the sum is over all states of the environment. After these operations, we have the output mixed state $\rho'$ from which we can calculate the transition probabilities and the mutual information between output and input under errors.

\subsection{Gate errors}

We model gate errors at the logical level with Pauli noise channels after each gate. For a single qubit:

\[A_0 = \sqrt{1-p_x-p_y-p_z} I\]

\[A_1 = \sqrt{p_x} X \]

\[A_2 = \sqrt{p_y} Y\]

\[A_3 = \sqrt{p_z} Z\]

\[\rho' = (1-p_x-p_y-p_z)\rho + p_x X \rho X +  p_y Y \rho Y +  p_z Z \rho Z\]

For a multiple qubit Pauli channel, the probabilities of each phase and/or bit flip error depend on the error model. We consider three different models:
independent, uniform and paired. The uniform and paired models are types of \emph{correlated} error models. In the main text we showed the results for the independent model. We here explain the different models and present results for the two not included in the main text (uniform and paired).

\subsubsection{Independent Pauli noise}
For a 2-qubit \emph{independent} Pauli channel, the phase and/or bit flip errors on each qubit are independent events. Therefore, we have the following expressions for the error probabilities (assuming the phase and/or bit flip error for both qubits is equal):

\begin{align*}
    \Pr(\mathbb{I} \otimes \mathbb{I}) =& (1-p_x-p_y-p_z)^2 \\
    \Pr(\mathbb{I} \otimes X) =& (1-p_x-p_y-p_z)(p_x) \\
    \Pr(\mathbb{I} \otimes Y) =& (1-p_x-p_y-p_z)(p_y)\\
    ... \\
    \Pr(Z \otimes Y) =& p_z p_y \\
    \Pr(Z \otimes Z) =& p_z^2  
\end{align*}

\noindent The receiver performance for independent Pauli noise is presented in the main text, Figure 5(b).

\subsubsection{Correlated Pauli noise}

For a general 2-qubit Pauli channel we have the following for the error probabilities: 

\begin{align*}
    \Pr(\mathbb{I} \otimes \mathbb{I}) =& p_{II}\\
    \Pr(\mathbb{I} \otimes X) =& p_{IX}\\
    \Pr(\mathbb{I} \otimes Y) =& p_{IY}  \\
    ...\\
    \Pr(Z \otimes Y) =& p_{ZY}\\
    \Pr(Z \otimes Z) =& p_{ZZ} 
\end{align*}

\noindent such that $p_{II} + p_{IX} + ... + p_{ZY} + p_{ZZ} = 1$

For the \emph{paired} model, an error on one qubit implies an error on both. We choose all such possible errors to occur with equal probability such that we have $p_{II} = p_{IX} = p_{IY} = ... = p_{ZI} = 0$ and all other $p_{AB}$ are the same. For the \emph{uniform} error model, each 2 qubit error type occurs with equal probability: $p_{II} = p_{IX} = ... = p_{ZY} = p_{ZZ}$. The performance of the receiver under these correlated error models is shown in Figure 4.

\begin{figure}[H]
    \centering
    \subfigure[]{
        \includegraphics[height=9cm, width=12cm]{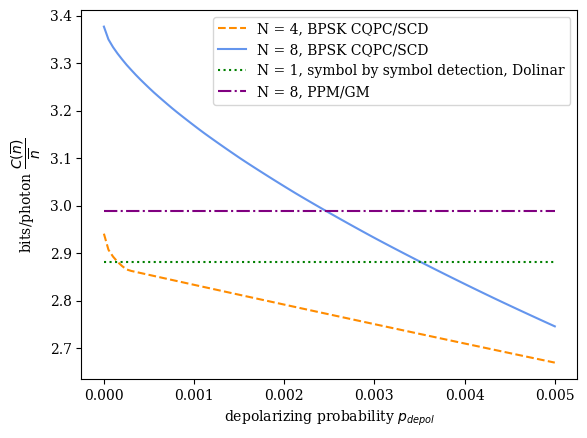}
        }
    \hspace{0.01cm}
    \subfigure[]{
        \includegraphics[height=9cm, width=12cm]{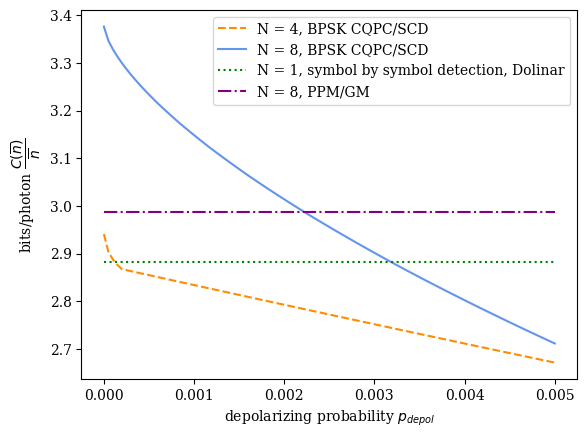}
    }
    \caption{Receiver performance with (a) uniform Pauli noise, (b) paired Pauli noise. }
\end{figure}

\subsubsection{Errors in the compression step}

Unlike gate errors in the decoding step, gate errors in the compression step also involve the logical control qubit. Since the control is reset between time bins, the PIE for errors in this step was calculated by treating the CNOT gates in each time bin as having independent control qubits. A circuit representation of the process and the inclusion of errors is shown below for a sample time-bin encoding, for $N=4$. The top 4 wires are for the controls, which are decoupled in each time bin, and the bottom 3 wires are for the target processor qubits that store the binary encoding of the input photonic state.

\bigskip
\Qcircuit @C=1em @R=2em {
&  \qw & \qw  & \qw & \qw & \qw & \qw & \qw & \qw & \qw & \qw & \qw & \ctrl{4} & \gate{\mathcal{N}} & \qw & \ctrl{5} & \gate{\mathcal{N}} & \ctrl{6} & \gate{\mathcal{N}} & \gate{H} & \gate{\mathcal{M}} & \meter\\
&  \qw & \qw & \qw & \qw & \qw & \qw & \qw & \ctrl{3}  & \gate{\mathcal{N}} & \ctrl{4} & \gate{\mathcal{N}} & \qw  & \qw & \qw & \qw & \qw & \qw & \qw & \gate{H} & \gate{\mathcal{M}} & \meter \\
&   \qw & \qw & \qw & \ctrl{2} & \gate{\mathcal{N}} & \ctrl{4} & \gate{\mathcal{N}} & \qw & \qw & \qw & \qw & \qw & \qw & \qw & \qw & \qw & \qw & \qw & \gate{H} & \gate{\mathcal{M}} & \meter\\
&   \qw & \ctrl{1} & \gate{\mathcal{N}} & \qw  & \qw & \qw & \qw & \qw & \qw & \qw & \qw & \qw & \qw & \qw & \qw & \qw & \qw & \qw &\gate{H} & \gate{\mathcal{M}} & \meter \\
&   \qw & \targ & \gate{\mathcal{N}} & \targ & \gate{\mathcal{N}} & \qw & \qw & \targ & \gate{\mathcal{N}} & \qw & \qw & \targ & \gate{\mathcal{N}} & \qw & \qw & \qw & \qw & \qw & \qw \\
&   \qw & \qw & \qw & \qw  & \qw & \qw & \qw & \qw & \qw & \targ & \gate{\mathcal{N}} & \qw & \qw & \qw & \targ & \gate{\mathcal{N}} & \qw & \qw & \qw \\
&   \qw & \qw & \qw & \qw & \qw  & \targ & \gate{\mathcal{N}} & \qw & \qw & \qw & \qw & \qw & \qw & \qw & \qw & \qw & \targ & \gate{\mathcal{N}} & \qw 
}
\bigskip

For the control qubit decoupling part of the compression step, we perform a Hadamard gate with Pauli noise, then project the control qubit state to the 0 state and trace out the control qubit. After that, the remaining target qubit state is fed into the decoding circuit. As in the main text, $\mathcal{M}$ blocks represent single qubit depolarizing noise, while $\mathcal{N}$ blocks represent 2-qubit depolarizing noise.

\subsection{Transducer errors}
Errors in the transduction step can be viewed as depolarizing errors on the input photonic state, $A \ket{0} + \alpha_i \ket{1}$, in each time-bin $i$ that ultimately lead to an incorrect recording of the state for that time-bin in the processor register.

With some error rate $\epsilon$ in each time-bin and with $N$ time-bins, we can model this as:

\[\ket{\psi_{reg}}\bra{\psi_{reg}} \rightarrow (1-\epsilon)^N \ket{\psi_{reg}}\bra{\psi_{reg}} + A\epsilon(1-\epsilon)^{N-1} \sum_{i=1}^N \frac{\mathbb{I}_i}{2} \bigotimes_{j=1,j \neq i}^N \ket{0}\bra{0}_j+ O((1-\epsilon)^{N-2}A\epsilon^2) + \sum_{i=1}^N O((1-\epsilon)^{N-1}\alpha_i \epsilon)\]

Here, $\ket{\psi_{reg}}$ is the ideal register state without errors, while the second term corresponds to single qubit depolarizing errors on the part of the register state corresponding to the vacuum. The third term corresponds to errors on two or more qubits in the part of the register state corresponding to the vacuum and the fourth term corresponds to errors in the single-photon terms.

In the regime we are focused on in this work, we have $A \sim 1$ and $\alpha_i \ll 1$. The dominant error term is thus the $A\epsilon(1-\epsilon)^{N-1}$, which represents single phase/bit flip errors in the vacuum state (\emph{dark count} errors). We thus model transducer errors as depolarizing noise on the input photonic state $\rho=\ket{\psi_{ph}}\bra{\psi_{ph}}$ ($\ket{\psi_{ph}} =  \ket{00} + \alpha[\ket{10}] + \ket{01}]$ for $N=2$), occurring with probability $p$:

\[\rho' = (1- p) \rho + \frac{p}{N+1}\mathbb{I} \]

Here, the subspace is truncated to the vacuum and single photons, an approximation which reflects the dominance of dark count errors.